\documentclass[letterpaper,british,english,reprint, pra]{revtex4-1}
\usepackage[T1]{fontenc}
\usepackage[latin9]{inputenc}
\usepackage{babel}
\usepackage{units}
\usepackage{textcomp}
\usepackage{amsmath}
\usepackage{amssymb}
\usepackage{graphicx}
\usepackage[unicode=true]
 {hyperref}

\makeatletter

\pdfpageheight\paperheight
\pdfpagewidth\paperwidth

\newcommand{\lyxmathsym}[1]{\ifmmode\begingroup\def\b@ld{bold}
  \text{\ifx\math@version\b@ld\bfseries\fi#1}\endgroup\else#1\fi}

\makeatother

\begin{document}

\preprint{APS/123-QED}

\title{An improved lower bound for superluminal quantum communications.}

\author{Bruno Cocciaro}
\email{b.cocciaro@comeg.it}

\affiliation{Department of Physics Enrico Fermi, Largo Pontecorvo 3, I-56127 Pisa,
Italy.}

\author{Sandro Faetti}
\email{sandro.faetti@unipi.it}

\affiliation{Department of Physics Enrico Fermi, Largo Pontecorvo 3, I-56127 Pisa,
Italy.}

\author{Leone Fronzoni}
\email{leone.fronzoni@unipi.it}

\affiliation{Department of Physics Enrico Fermi, Largo Pontecorvo 3, I-56127 Pisa,
Italy.}

\date{\today}
\begin{abstract}
Superluminal communications have been proposed to solve the Einstein,
Podolsky and Rosen (\emph{EPR}) paradox. So far, no evidence for these
superluminal communications has been obtained and only lower bounds
for the superluminal velocities have been established. In this paper
we describe an improved experiment that increases by about two orders
of magnitude the maximum detectable superluminal velocities. The locality,
the freedom-of-choice and the detection loopholes are not addressed
here. No evidence for superluminal communications has been found and
a new higher lower bound for their velocities has been established.
\begin{description}
\item [{PACS~numbers}] 03.65.Ud, 03.67.Mn{\small \par}
\end{description}
\end{abstract}



\maketitle

\section*{Introduction}

\begin{figure}[h]
\centering{}\includegraphics[scale=0.3]{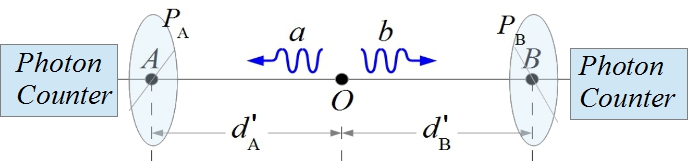}\caption{\label{fig:wide-1-1}Two entangled photons \emph{a} and \emph{b} are
generated at \emph{O} and get the absorption polarizing films $P_{A}$
(Alice polarizer) and $P_{B}$(Bob polarizer). The photons passing
through the polarizers are collected by photon counting modules. With
$d'_{A}$ and $d'_{B}$ we denote the optical paths of photons \emph{a}
and \emph{b} from source \emph{O }to polarizers $P_{A}$ and $P_{B}$,
respectively.}
\end{figure}

In 1935 Einstein, Podolsky and Rosen\citep{EPR} showed that orthodox
Quantum Mechanics (\emph{QM}) is a non-local theory (\emph{EPR} paradox).
Consider, for instance, photons \emph{a} and \emph{b} in Figure \ref{fig:wide-1-1}
that propagate in opposite directions and that are in the polarization
entangled state

\begin{equation}
\left|\psi\right\rangle =\frac{1}{\sqrt{2}}\left(\left|H,H\right\rangle +\left|V,V\right\rangle \right)\quad,\label{eq:1}
\end{equation}
where \emph{H} and \emph{V} denote horizontal and vertical polarization,
respectively. According to \emph{QM}, a polarization measurement on
photon \emph{a} leads to the instantaneous collapse of the polarization
state of photon \emph{b} whatever is its distance from \emph{a}. This
behavior is reminiscent of the action at a distance that has been
completely rejected by the General Relativity and the Electromagnetism
theories. For this reason, Einstein et al. believed that \emph{QM}
is a not complete theory and suggested that a complete theory should
contain some additive local variables. In 1961 J. Bell showed \citep{Bell}
that any theory based on local variables must satisfy an inequality
that is violated by \emph{QM}.

Analogous inequalities have been found by Clauser et al. \citep{Clauser1_PhysRevLett1969,Clauser2_PhysRevD1974}.
The Aspect experiment of 1982 \citep{Aspect} demonstrated that the
Bell inequality is not satisfied and also showed that quantum correlations
cannot be explained in terms of subluminal or luminal communications.
Many other experiments confirmed the Aspect results and some recent
experiments finally closed the residual loopholes \citep{Loophole1,Loophole2,Loophole3,Loophole4}.
Then, the experimental results demonstrate that the local variables
models cannot explain the quantum correlations between entangled particles.
Some physicists suggested \citep{Eberhard_1989,Bohm_undivided_1993}
that these correlations could be due to superluminal communications
\footnote{The key idea is that, in a typical \emph{EPR} experiment, the two
measurements are not exactly simultaneous. When the first measurement
is performed in the point \emph{A}, a collapsing wave propagates superluminally
in the space starting from \emph{A}. The predicted \emph{QM} correlations
(e. g. the violation of the Bell inequality) occur only if the collapsing
wave gets the second particle before its measurement.} (\emph{v}-causal models in nowadays literature \citep{Gisin2014}).
To avoid causal paradoxes, they assumed that a preferred frame (\emph{PF})
exists where superluminal signals propagate isotropically with unknown
velocity $v_{t}=\beta_{t}\:c$ ($\beta_{t}>1$). Below we will indicate
the relativistic parameter $\beta_{t}=v_{t}/c$ as ``the adimensional
velocity''. Someone could be surprised for the existence of a preferred
frame but references \citep{Cocciaro_2013_ShutYourselfUp,Cocciaro3_DICE2015}
strongly stressed that the existence of a \emph{PF} is not in the
contrast with relativity. Furthermore, it has to be noticed that an
universal \emph{PF} has been already observed: it is the Cosmic Microwave
Background frame (\emph{CMB frame)} that moves at the adimensional
velocity $\beta\approx10^{-3}$ with respect to the Earth frame. It
has been recently demonstrated an important theorem \citep{Bancal_NatPhys_2012,Barnea_PhysRevA2013}:
\emph{v}-causal models allow superluminal communications in the macroscopic
world (signalling) if more than 2 entangled particles are involved.
 Although one of us believes that signalling is not incompatible with
relativity \citep{Cocciaro_2013_ShutYourselfUp,Cocciaro3_DICE2015},
most physicists think that there is no compatibility and that the
experimental evidence of signalling would need a revision of relativity.
In standard conditions, the superluminal communications lead to the
usual \emph{QM} correlations but there are special conditions (if
the second particle reaches its measurement device when the collapsing
wave didn't yet reach it) where the \emph{QM} correlations cannot
be established and the Bell inequality should be satisfied. In fact,
if the absorption polarizing films $P_{A}$ and $P_{B}$ in Figure
\ref{fig:wide-1-1} are at the same optical paths $d'_{A}$ and $d'_{B}$
from source \emph{O} in the \emph{PF}, the two photons get them simultaneously
and there is no time to establish \emph{QM} correlations. To verify
this behavior, one can measure the correlation parameter $S_{max}$
defined as \citep{Aspect_2002,Cocciaro2_DICE2016} 

\begin{equation}
S_{max}=P_{0}-\sum_{i=1}^{3}P_{i}\;,\label{eq:Smax}
\end{equation}
with $P_{0}=P\left(45\text{\textdegree},67.5\text{\textdegree}\right)$,
$P_{1}=P\left(0\text{\textdegree},67.5\text{\textdegree}\right)$,
$P_{2}=P\left(45\text{\textdegree},112.5\text{\textdegree}\right)$
and $P_{3}=P\left(90\text{\textdegree},22.5\text{\textdegree}\right)$,
where $P(\alpha,\xi)$ is the probability that photon \emph{a} passes
through polarizer $P_{A}$ aligned at the angle $\alpha$ with respect
to the horizontal plane and that photon \emph{b} passes through polarizer
$P_{B}$ aligned at the angle $\xi$. For any local variables model,
$S_{max}$ must satisfy the modified Bell-Clauser-Horne-Shimony-Holt
inequality $S_{max}\leq0$ \citep{Aspect_2002,Cocciaro2_DICE2016}
whilst \emph{QM} predicts $S_{max}=\nicefrac{(\sqrt{2}-1)}{2}\approx0.2071$
for the entangled state in eq.(\ref{eq:1}). Probabilities $P(\alpha,\xi)$
can be experimentally obtained using the relation

\begin{equation}
P(\alpha,\xi)=\frac{N\left(\alpha,\xi\right)}{N_{tot}}\;,\label{eq:Smax-1}
\end{equation}
where $N(\alpha,\xi)$ are the coincidences between entangled photons
passing through the polarizers during the acquisition time $\Delta t$
and $N_{tot}$ is the total number of entangled photons couples that
can be obtained using eq.(\ref{eq:NTOT}): 

\begin{equation}
N_{tot}=\sum_{i=0}^{3}N_{i}\;,\label{eq:NTOT}
\end{equation}
where $N_{0}=N\left(0\text{\textdegree},0\text{\textdegree}\right)$,
$N_{1}=N\left(0\text{\textdegree},90\text{\textdegree}\right)$, $N_{2}=N\left(90\text{\textdegree},0\text{\textdegree}\right)$
and $N_{3}=N\left(90\text{\textdegree},90\text{\textdegree}\right)$.
If $d\lyxmathsym{\textquoteright}_{A}=d\lyxmathsym{\textquoteright}_{B}$
in the \emph{PF}, the quantum correlations cannot be established and
$S_{max}$ should always satisfy the inequality $S_{max}\leq0$ \citep{Clauser1_PhysRevLett1969,Clauser2_PhysRevD1974,Aspect_2002,Cocciaro2_DICE2016}.
Due to the experimental uncertainty $\Delta d\lyxmathsym{\textquoteright}$
on the equalization of the optical paths in the \emph{PF}, the arrival
times of the entangled photons at the polarizers could differ from
one another for the quantity $\Delta t'=\nicefrac{\Delta d'}{c}$
and, thus, a superluminal communication would be impossible only if
$\Delta t'$ is lower than the communication time $d'_{AB}/(\beta_{t}c)$,
where $d'_{AB}$ is the optical path from \emph{A} to \emph{B} in
the \emph{PF} (see Figure 1). The above condition is satisfied only
if $\beta_{t}$ is lower than the maximum detectable adimensional
velocity $\beta_{t,max}=\nicefrac{d'_{AB}}{\Delta d'}$ of the superluminal
communications. Therefore, due to the $\Delta d\lyxmathsym{\textquoteright}$
uncertainty, a breakdown of quantum correlations ($S_{max}<0$) could
be observed only if $\beta_{t}<\beta_{t,max}$. In the Earth frame
the analysis becomes more complex. Indeed, the equalization of the
optical paths $d_{A}$ and $d_{B}$ in the Earth frame does not imply
their equalization also in the \emph{PF} except if the unknown adimensional
velocity vector $\overrightarrow{\beta}$ of the \emph{PF} with respect
to the Earth frame satisfies the orthogonality condition $\overrightarrow{\beta}\cdot\overrightarrow{AB}=0$.
If the \emph{AB} segment is East-West aligned, due to the Earth rotation
around its axis, there are always two times $t_{1}$ and $t_{2}$
for each sidereal day where $\overrightarrow{AB}$ becomes orthogonal
to $\overrightarrow{\beta}$ whatever is the orientation of the $\overrightarrow{\beta}$
vector \citep{Salart_nature_2008,Cocciaro_PLA_2011}. At these times,
the quantum correlations should not be established and $S_{max}$
should exhibit a breakdown from the quantum value $S_{max}=0.2071$
toward $S_{max}\leq0$ if the superluminal adimensional velocity $\beta_{t}$
is lower than $\beta_{t,max}$. However, an acquisition time $\Delta t$
has to be spent to measure parameter $S_{max}$ and, thus, the orthogonality
condition $\overrightarrow{\beta}\cdot\overrightarrow{AB}=0$ can
be only approximately satisfied during this acquisition time. This
leads to a further contribution to the uncertainty $\Delta t'$ on
the arrival times of the entangled photons at the two polarizers in
the preferred frame. Then, in the Earth experiment, the maximum detectable
velocity $\beta_{t,max}$ is affected both by the uncertainty $\Delta d$
on the equalization of the optical paths and by the acquisition time
$\Delta t$. Smaller ones are $\Delta d$ and $\Delta t$ and bigger
is $\beta_{t,max}$. Using the relativistic Lorentz equations one
finds \citep{Salart_nature_2008,Cocciaro_PLA_2011}

\begin{equation}
\beta_{t,max}=\sqrt{1+\frac{\left(1-\beta^{2}\right)\left[1-\rho^{2}\right]}{\left[\rho+\frac{\pi\beta\delta t}{T}\sin\chi\right]^{2}}}\;,\label{eq:betamin-2}
\end{equation}
where $\chi$ is the unknown angle that the velocity of the \emph{PF}
makes with the Earth rotation axis, \emph{T} is Earth rotation day
and $\rho=\nicefrac{d_{AB}}{\Delta d}$ , where $d_{AB}$ is the optical
path between points \emph{A} and \emph{B} in the Earth Frame. Parameter
$\delta t$ ($\delta t/T\ll1$) in eq.(\ref{eq:betamin-2}) has been
usually assumed to coincide with time $\Delta t$ needed for a complete
measurement of $S_{max}$ but this is not correct. Indeed, if $t_{i}$
(\emph{i} =1,2) are the daily times where the orthogonality condition
$\overrightarrow{\beta}\cdot\overrightarrow{AB}=0$ is satisfied,
the superluminal model predicts that no communication is possible
in the time intervals $I_{i}=\left[t_{i}-\delta t/2,\:t_{i}+\delta t/2\right]$
if $\beta_{t}<\beta_{t,max}$ \citep{Salart_nature_2008,Cocciaro_PLA_2011}.
Unfortunately, times $t_{i}$ are unknown and the acquisitions cannot
be synchronized with them. Then, one can be sure that a full acquisition
interval $\Delta t$ is certainly contained in the unknown $I_{i}$
interval only if $\Delta t\leq\delta t/2$. This means that parameter
$\delta t$ in eq.(\ref{eq:betamin-2}) is given by

\begin{equation}
\delta t=2\,\Delta t\;.\label{eq:betamin-2-1}
\end{equation}

Some experimental tests of the superluminal models have been reported
in the literature but, so far, no evidence for a violation of \emph{QM}
predictions has been found and only lower bounds $\beta_{t,max}$
have been established \citep{Salart_nature_2008,Cocciaro_PLA_2011,Cinesi_PhysRevLett2013,Cocciaro2_DICE2016}.
In reference \citep{Cinesi_PhysRevLett2013} the locality and freedom-of-choice
loopholes were also addressed. Here we report the results of a new
experiment where the loopholes above are not taken into account but
the maximum detectable velocity of the superluminal communications
is increased by about two orders of magnitude. In particular, according
to \citep{SalartReply} we here test the ``assumption that quantum
correlations are due to supra-luminal influences of a first event
onto a second event''. Since we use absorption polarizing films,
we assume that the above events are the collapses of the polarization
state that occur when photons hit the absorption polarizers.

\section{EXPERIMENTAL METHOD}

\subsection{The experimental apparatus and procedures.}

Our experimental apparatus, the procedures used to get very small
values of the basic experimental parameters $\rho$ and $\Delta t$
and the experimental uncertainties have been described in detail in
a previous paper \citep{Cocciaro2_DICE2016} and, thus, we will remind
here only the main features. 

\begin{figure*}[t]
\begin{centering}
\includegraphics[scale=0.6]{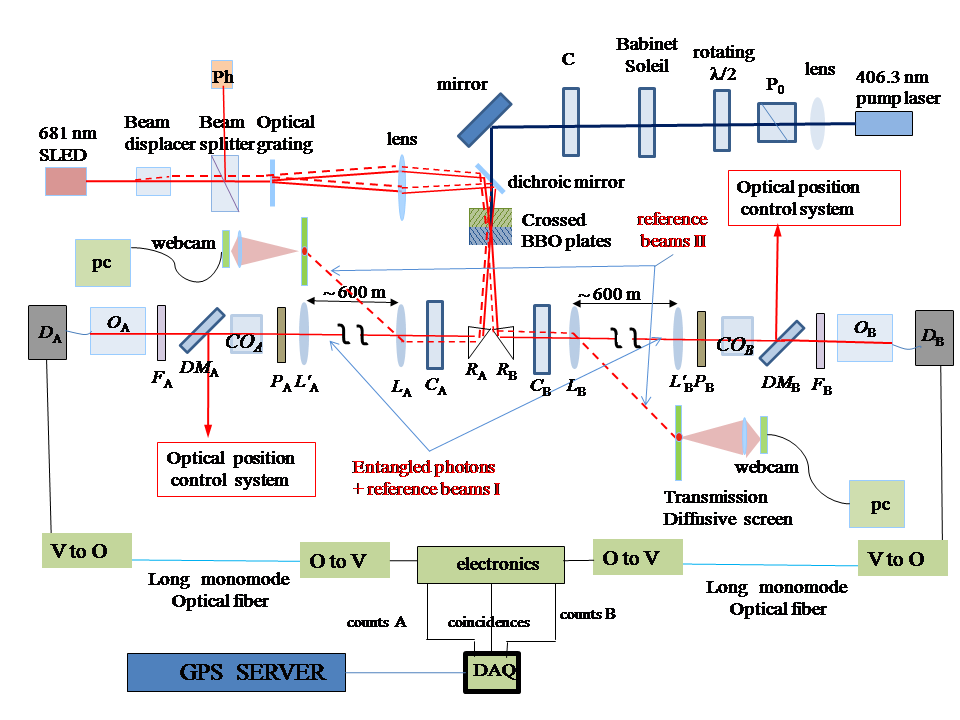}
\par\end{centering}
\caption{\label{fig:wide-1}\foreignlanguage{british}{ Schematic view of the
experimental apparatus. Note that the figure is not to scale and,
in particular, the distances between lenses $L_{A}\thinspace\mathrm{and}\thinspace L_{B}$
and $\thinspace L'_{A}$ and $L'_{B}$ ($\approx600\thinspace\mathrm{m}$)
are much larger then all the other distances. To simplify the drawing
some details have not be inserted in the figure. The 220 mW pump beam
with wavelength $\lambda_{p}=$406.3 nm (blue thick full line in the
figure) is polarized by the polarizer $P_{0}$ and the $\lambda/2$
plate. The Babinet-Soleil compensator introduces a variable optical
dephasing between the horizontal and vertical polarizations. $C$,
$C_{A}$ and $C_{B}$ are anisotropic compensator plates used to get
a high intensity source of entangled photons with a sufficient fidelity.
$R_{A}$ and $R_{B}$ are right angle prisms. The pump beam is focused
at the centre of two crossed adjacent \emph{BBO} plates (29.05\textdegree{}
tilt angle) where entangled photons having wavelength $\lambda=$812.6
nm are generated and emitted at the angles $\pm2.42\text{\textdegree}$
with respect to the pump laser beam. $L_{A},\thinspace L_{B},\thinspace L'_{A}$
and $L'_{B}$ are specially designed $15\thinspace\text{cm}$-diameter
achromatic lenses aligned along the \emph{EGO} gallery and having
a $6.00\thinspace\text{m}$ focal length at both the 812.6 nm and
681 nm wavelengths. $P_{A}$ and $P_{B}$ are absorption polarizing
filters. $O_{A}$, $O_{B}$, $CO_{A}$ and $CO_{B}$ are systems of
lenses. $DM_{A}$ and $DM_{B}$ are dichroic mirrors, $F_{A}$ and
$F_{B}$ are sets of adjacent optical filters, $D_{A}$ and $D_{B}$
are photon counting detectors. The superluminous diode (\emph{SLED})
having wavelength $\lambda_{R}$= 681 nm and coherence length 28.1
${\textstyle \mu m}$, the beam displacer and the optical grating
are used to produce two reference beams in each arm of the \emph{EGO}
gallery (full and broken red lines) as discussed in the text. \emph{V}
to \emph{O} denote electronic systems that transform the output voltage
pulses produced by the photon counting detectors into optical pulses,
whilst \emph{O} to \emph{V} transform the optical pulses into voltage
pulses. \emph{DAQ} is a National Instruments CompactDAC that provides
a real time acquisition of coincidences.}}
\end{figure*}
To reach a high value of $\beta_{t,max}$, one has to make parameters
$\rho$ and $\delta t$ in eq.(\ref{eq:betamin-2}) as smaller as
possible. We get a small value of $\rho=\nicefrac{d_{AB}}{\Delta d}$
performing our measurements in the so called \textquotedbl{}East-West\textquotedbl{}
gallery of the European Gravitational Observatory (\href{https://www.ego-gw.it/}{EGO})\citep{EGO}
of Cascina ($d_{AB}\approx1200$ m) and we use an interferometric
method to equalize the optical paths $d_{A}$ and $d_{B}$ ($d_{A}\approx d_{B}\approx600$
m). The final uncertainty $\Delta d$ on the equality of the optical
paths is due to many error sources including the finite thickness
of the polarizing layers, the air dispersion and the uncertainty on
the interferometric measurement. As shown in reference \citep{Cocciaro2_DICE2016},
the estimated uncertainty is $\Delta d\approx0.22$ mm. To reduce
the acquisition time we need a high intensity source of entangled
photons in a sufficiently pure entangled state. We get this goal using
the compensation procedures developed by the Kwiat group \citep{Kwiat_OptExpr_2005,Kwiat_OptExpr_2007,Kwiat_OptExpr_2009}
and developing a proper optical configuration that ensures low losses
of entangled photons along the gallery. Unfortunately, the \emph{EGO}
gallery is not aligned along the East-West axis but makes the angle
$\gamma=18\text{\textdegree\ }$ $=\pi/10$ with it. Then, one easily
infers that the orthogonality condition $\overrightarrow{\beta}\cdot\overrightarrow{AB}=0$
can be never satisfied if the velocity vector of the \emph{PF} makes
a polar angle $\chi<\gamma=\pi/10$ or $\chi>\pi-\gamma=9\pi/10$
with respect to the Earth rotation axis. This means that our experiment
is virtually insensitive to a fraction \foreignlanguage{british}{$\Omega/(4\,\pi)=\int_{_{0}}^{\gamma}\sin\theta d\theta<5\%$}
of all the possible alignments of the \emph{PF} velocity vector. For
a detailed analysis of the case $\gamma\neq0$ we refer the reader
to reference \citep{Salart_nature_2008}. Note that the Reference
Frame of the Cosmic Microwave Background (\emph{$\chi\approx97\text{\textdegree}$})
is accessible to our experiment. Eq.(\ref{eq:betamin-2}) was obtained
under the assumption that the experiment is aligned along the East-West
axis ($\gamma=0$) but, for $\gamma\geq0$ and $\pi-\gamma\geq\chi\geq\gamma$,
it has to be replaced by \citep{Salart_nature_2008} 

\begin{equation}
\beta_{t,max}=\sqrt{1+\frac{\left(1-\beta^{2}\right)\left[1-\rho^{2}\right]}{\left[\rho+A\,\frac{\pi\beta\delta t}{T}\right]^{2}}}\;,\label{eq:betamin-2-2}
\end{equation}
where coefficient \emph{A} is defined as 

\begin{equation}
A=\sqrt{\sin^{2}\chi\cos^{2}\gamma-\cos^{2}\chi\sin^{2}\gamma}\:.\label{eq:betamin-2-2-1}
\end{equation}
Velocity $\beta_{t,max}$ greatly decreases out of the interval $\pi-\gamma\geq\chi\geq\gamma$\citep{Salart_nature_2008}.

A schematic view of the experimental apparatus is shown in figure
\ref{fig:wide-1}. A pump laser beam at a wavelength $\lambda_{p}=$406.3
nm is generated by the 220 mW laser diode shown at the top right in
figure \ref{fig:wide-1}. The pump beam passes through an achromatic
lens, a Glan-Thompson polarizer, a motorized $\lambda/2$ plate, a
motorized Babinet-Soleil compensator and a quartz plate \emph{C}.
Then, it is reflected by a mirror, passes through a 565 nm short pass
dichroic mirror (Chroma T565spxe) and is focused (spot diameter =
0.6 mm) at the center of two thin (thickness $\approx0.56$ mm) adjacent
crossed\textbf{ }\emph{BBO} nonlinear optical crystals plates (29.05\textdegree{}
tilt angle) cut for \emph{type I phase matching} \citep{Kwiat_PhysRevA_1999}.
The \emph{BBO} plates have the optical axes lying in the horizontal
and vertical plane, respectively. The $\lambda/2$ plate aligns the
polarization of the incident pump beam at 45\textdegree{} with respect
to the horizontal axis. The quartz plate \emph{C} compensates the
effects due to the low coherence of the pump beam ($\approx$ 0.2
mm coherence length) \citep{Kwiat_OptExpr_2009}. Down conversion
leads to two outgoing beams of entangled photons at the average wavelength
$\lambda=2\,\lambda_{p}=812.6\,\textrm{nm}$ that mainly propagate
at two symmetric angles ($\pm2.4\text{2\textdegree}$) with respect
to the normal to the crossed \emph{BBO} plates. A proper adjustment
of the optical dephasing induced by the Soleil-Babinet compensator
provides the polarization entangled state in eq.(\ref{eq:1}). The
entangled beams are deviated in opposite directions along the \emph{EGO}
gallery by two right-angle prisms ($R_{A}$ and $R_{B}$) and passe
through the \emph{BBO} compensating plates $C_{A}$ and $C_{B}$.
The Kwiat compensating plates \emph{C}, $C_{A}$ and $C_{B}$ are
used to get a high intensity source of entangled photons ($N_{tot}\approx23000$
coincidences/s) in an entangled state of sufficient purity \citep{Kwiat_OptExpr_2005,Kwiat_OptExpr_2007,Kwiat_OptExpr_2009}.
The entangled beams, propagating along opposite directions, impinge
on polarizers $P_{A}$ and $P_{B}$ at a distance of about 600 m from
the source. Our experiment requires the equalization of the optical
paths $d_{A}$ and $d_{B}$ between the source of the entangled photons
and polarizers $P_{A}$ and $P_{B}$ and needs stable coincidences
counts during the whole measurement time ($\approx$ 8 days). Both
these requirements are satisfied using four reference beams at wavelength
$\lambda_{R}=$681 nm that are utilized to align the optical system,
to equalize the optical paths and to compensate the deviations of
the entangled beams due to the air refractive index gradients induced
by sunlight on the top of the gallery. The four reference beams are
obtained starting from the collimated beam emitted by the 3 mW superluminous
diode (\emph{SLED}) shown at the top left in figure \ref{fig:wide-1}.
The beam passes through a beam displacer (Thorlabs BDY12U) that splits
the incident beam into two parallel beams (I and II) at a relative
distance of 1.2 mm. Beam I is represented by a full line in the figure
whilst beam II by a broken line. Beams I and II are focused (spot
diameter $\approx0.3$ mm) orthogonally on a transmission phase grating
that mainly produces +1 e -1 diffracted beams at the diffraction angles
$\pm$ 2.43\textdegree{} that are virtually coincident with the average
emission angles of the entangled photons ($\pm$2.42\textdegree ).
An achromatic lens (150 mm focal length) projects on the crossed \emph{BBO}
plates a 1:1 image of the spots of beams I and II occurring on the
grating. The spot of beam I is centered within $\approx\pm0.03\:{\textstyle mm}$
with respect to the pump beam spot where the entangled photons are
generated (the ``source'' of the entangled photons). Then, beams
I outgoing from the crossed \emph{BBO} plates virtually follows the
same paths of the entangled beams. The whole system described above
lies on an optical table and is enclosed in a large box that ensures
a fixed temperature $T=24\text{\textdegree C}\pm0.1\text{\textdegree C}$
by circulation of Para-flu fluid. Two 80 W fans ensure a sufficient
temperature uniformity. The entangled beams and the reference beams
are collected by large diameter (15 cm) achromatic lenses $L_{A}$
and $L_{B}$ that have been built to have the same focal length at
the wavelengths of the reference and the entangled beams (6.00 m at
$\lambda_{R}=681\,\textrm{nm and }$$\lambda=812.6\,\textrm{nm}$).
These beams propagate along the gallery arms and impinge on two identical
achromatic lenses $L'_{A}$ and $L'_{B}$ at a distance $\approx$
600 m from the source of the entangled photons. Real 1:1 images (0.6
mm-width) of the source and of the spot of beam I occurring on the
crossed \emph{BBO} plates are produced at the centers of the linear
polarizers layers $P_{A}$ and $P_{B}$ (Thorlabs \emph{LPNIR }).
Beams II are slightly deviated by lenses $L_{A}$ and $L_{B}$ and
impinge on two diffusing screens put adjacent to lenses $L'_{A}$
and $L'_{B}$. The diffused light outgoing from each screen is collected
by a webcam connected to a \emph{PC} and a Labview program calculates
the position of the diffusing spot. The daily displacements of the
above spots (up to 1.2 m in a Summer day) due to air refractive index
gradients induced by sunlight are compensated using a proper feedback
where lenses $L_{A}$ and $L_{B}$ are moved orthogonally to their
optical axes to maintain fixed the position of the spots on the diffusing
screens (see Section 2.2 in reference \citep{Cocciaro2_DICE2016}
for details). This procedure ensures that beams I and, thus, the entangled
beams remain virtually centered with respect to lenses $L'_{A}$ and
$L'_{B}$. The reference beams I outgoing from polarizers $P_{A}$
and $P_{B}$ are almost fully reflected by the long pass dichroic
mirrors $DM_{A}$ and $DM_{B}$ (Chroma T760lpxr) and enter the optical
position control systems that measure the position and the astigmatism
of the beam spots on the polarizers. Using a Labview program operating
in a \emph{PC}, lenses $L'_{A}$ and $L'_{B}$ are moved orthogonally
to their optical axes to maintain the spot position at the center
of the polarizers within $\pm0.4$ mm during the whole measurement
time. An other program controls the astigmatism of the images using
the variable-focus cylindrical lenses $CO_{A}$ and $CO_{B}$. The
equalization of the optical paths $d_{A}$ and $d_{B}$ is obtained
exploiting the beams I that are partially reflected by the polarizing
layers $P_{A}$ and $P_{B}$ and that come back producing interference
on the photodetector \emph{Ph} shown on the top left in Figure \ref{fig:wide-1}.
Details on the feedback procedures and on the interferometric method
can be found in Section 2.2 and 2.3 of reference \citep{Cocciaro2_DICE2016},
respectively. Each of the entangled photons beams outgoing from the
two polarizers passes through the long pass dichroic mirror ($DM_{A}$
or $DM_{B}$ in the Figure) and a filtering set ($F_{A}$ or $F_{B}$
in the Figure) made by two long-pass optical filters ( \foreignlanguage{british}{Chroma
ET765lp filters ; $\lambda_{c}=765\,\mathrm{nm}$)} that stop the
reference 681 nm beams and a band-pass filter \foreignlanguage{british}{(
Chroma ET810/40m ; $\lambda=810\,\mathrm{nm}\pm20\,\mathrm{nm}$)}.
Then, each beam is focused by a system of optical lenses ($O_{A}$
or $O_{B}$ ) on a $200\,\mu m$ multimode optical fiber having a
large numerical aperture (0.39) connected to a Perkin Elmer photons
counter module. The output pulses of the photons counters are transformed
into optical pulses (using the LCM155EW4932-64 modules of Nortel Networks)
that propagate in two monomode optical fibers toward the central optical
table where the entangled photons are generated. Finally, the optical
pulses are transformed again into electric pulses and sent to an electronic
coincidence circuit. An electronic counter connected to a National
Instruments CompactDAQ counts the Alice pulses $N_{A}$, the Bob pulses
$N_{B}$ and the coincidences pulses \emph{N}. 

\subsection*{B. The fast acquisition procedure.}

In our preliminary experiment \citep{Cocciaro2_DICE2016}, the measurements
of the probabilities appearing in eq.(\ref{eq:Smax}) were made sequentially:
a \emph{PC} connected to precision stepper motors rotated polarizers
$P_{A}$ and $P_{B}$ up to reach the first couple of angles $\alpha$
and $\beta$ appearing in eq.(\ref{eq:Smax}) ($\alpha$ = 45\textdegree{}
and $\beta$= 67.5\textdegree ) and the corresponding coincidences
$N(\alpha,\beta)$ were acquired with an acquisition time of 1 s,
then the successive couple of $\alpha$ and $\beta$ angles was set
and the corresponding coincidences were acquired and so on. When all
the eight contributions $N(\alpha,\beta)$ entering in equations (\ref{eq:Smax})
and (\ref{eq:NTOT}) were obtained, the program calculated $S_{max}$.
This procedure needed many consecutive rotations of the polarizers
before a single value of $S_{max}$ was obtained leading to a long
acquisition time interval $\Delta t\approx$100 s for each measurement
of $S_{max}$. To greatly reduce $\Delta t$ and increase the maximum
detectable adimensional velocity $\beta_{t,max}$, we exploits here
the daily periodicity of the investigated phenomenon and we measure
each of the four contributions appearing in eq.(\ref{eq:Smax}) in
successive daily experimental runs. This procedure allows us to set
the polarization angles $\alpha$ and $\xi$ only one time each day
before starting the measurement of $P_{i}$. Then, any retardation
due to the polarizers rotation is avoided. Furthermore the \emph{PC
}used in our previous experiment has been replaced here by a National
Instruments CompactDAQ where a Real Time Labview program runs. This
new procedure ensures a full continuity of the acquisitions and a
constant acquisition time. The obtained experimental values of the
basic parameters $\rho$ (see \citep{Cocciaro2_DICE2016}) and $\delta t$
appearing in eq.(\ref{eq:betamin-2}) are

\begin{equation}
\rho=1.83\times10^{-7}\;\;and\;\;\delta t=2\,\Delta t=0.494\,\textrm{s}\label{eq:betamin-2-2-1-1}
\end{equation}
that provide a $\beta_{t,max}$ value about two orders of magnitude
higher than the those obtained in previous experiments. A \emph{GPS}
Network Time Server (TM2000A) provides the actual \emph{UTC} time
\citep{UTC,UT1time} with an absolute accuracy better than 1 ms also
if the connection to the satellites is lost up to a 80 hours time.
Since the investigated phenomenon is related to the Earth rotation,
we synchronize the acquisitions with the Earth rotation time $t=\theta\times240\,\textrm{s}$
where $\theta$ is the Earth Rotation Angle(\emph{ERA} \citep{UTC,UT1time})
expressed in degrees. The \emph{ERA} time is the modern alternative
to the Sidereal Time and it is given by $t=86400\times(\mathrm{TJ\:mod}\:1)$
where ``mod'' represents the modulo operation and $TJ=\left[a_{1}+b_{1}\times\left(\textrm{Julian\,UT1\,day}-2451545.0\right)\right]$
with $a_{1}=0.7790572732640$ days and $b_{1}=1.00273781191135448$.
The Julian UT1 day is strictly related to the \emph{UT1} time that
takes into account for the non uniformity of the Earth rotation velocity
and, thus, does not coincide with the \emph{UTC} atomic time provided
by the \emph{GPS}. The \href{https://datacenter.iers.org/eop/-/somos/5Rgv/latest/6}{IERS Bullettin A}
\citep{IERS} provides the value of the daily difference $\Delta=UT1-UTC$
and, thus, the\emph{ UT1} and the \emph{ERA} time can be calculated.
We decide to start each acquisition run at the Greenwich ERA time
$t=0$. 

The successive steps of the fully automated procedure are: 

\textbf{1}- The \emph{GPS} Greenwich \emph{UTC} time and the \emph{UT1}-\emph{UTC}
value are acquired, then, the Greenwich \emph{ERA} time \emph{t} is
calculated. Successively, the \emph{UTC} time that corresponds to
the next zero value of the Greenwich \emph{ERA} time is calculated. 

\textbf{2}- Two hours before the occurrence of $t=0$, we measure
the total number of couples of entangled photons $N_{tot}$. The program
rotates the $P_{A}$ and $P_{B}$ polarizers and sets successively
the $\alpha$ and $\xi$ angles that enter the expression of the total
number of incident entangled couples $N_{tot}$ in eq.(\ref{eq:NTOT}).
For each setting of the polarizers angles, the coincidences are measured
for a sufficiently long acquisition time interval (100 s) to made
negligible the counts statistical noise with respect to others noise
sources. The spurious statistical coincidences $N_{S}=N_{A}\times N_{B}\times T_{p}/\Delta t$
are subtracted, where $T_{p}$ is the pulses duration time and $\Delta t$
is the acquisition time interval. The value $T_{p}=$ 29.2 ns is obtained
from a calibration procedure where coincidences between totally uncorrelated
photons are detected. Finally, the total number of entangled photons
$N_{tot}$ is calculated using eq.(\ref{eq:NTOT}). 

\textbf{3}- At the end of these preliminary measurements, the polarizers
angles are set at the values $\alpha$= 45\textdegree{} and $\xi$=67.5\textdegree{}
appearing in the first contribution $P_{0}$ in eq.(\ref{eq:Smax}).
Then, the acquisition of the coincidences starts at the Greenwich
\emph{ERA} time $t=$0. The duration of a complete acquisition run
is $T_{0}=$ 36 \emph{ERA} hours that correspond to about 35 h ; 54
min and 7 s in the standard $UTC$ time. $2^{19}$ successive acquisitions
are made in each acquisition run with the acquisition time interval
$\Delta t=\nicefrac{T_{0}}{2^{19}}\simeq246.517461\,\textrm{ms}$
( in standard $UTC$ unities). Note that, due to the daily small changes
of the $UT1-UTC$ difference, $\Delta t$ exhibits small daily variations
(the maximum variation was $\approx$ 0.000001 ms in the whole measurement
time). To ensure a time precision better than 1 ms, the microseconds
internal counter of the Real Time Labview is used and the \emph{GPS}
server is interrogated every 5 minutes. Furthermore, a suitable subroutine
partially correct ( within 0.1 ms) time errors introduced by the microseconds
quantization of the \emph{DAQ} clock.\footnote{Due to the microseconds quantization of the $DAQ$ clock, the acquisition
time interval $\Delta t$ = 246517 ${\textstyle \mu s}$ is smaller
than $T_{0}/2^{19}=246517.46170157..{\textstyle \mu s}.$ by the quantity
$\Delta_{q}t$ = 0.46170157..$\mu s$. Then, the i-th acquisition
interval is shifted by (i-1)$\times\Delta_{q}t$ with respect to the
correct value (i-1)$\times\Delta t$ . As soon as this shift becomes
greater than 100 ${\textstyle \mu s}$ ( for a given $i$), the Labview
program increases the acquisition time of the i-th interval to $\Delta t+100\,{\textstyle {\displaystyle {\textstyle \mu s}}}$.
The same procedure is repeated whenever the successive shifts just
exceeds the 100 ${\textstyle \mu s}$ value. In such a way the maximum
residual shifts are always lower than $\approx100\,{\textstyle {\displaystyle {\textstyle \mu s}}}$
and, thus, are negligible with respect to the width $\Delta t$ =
246517 ${\textstyle \mu s}$ of each acquisition interval.} 

\textbf{4}- At the end of the first acquisition run, the program calculates
the $2^{19}$ values of $P_{0}$ and sets the second couple of angles
$\alpha$ and $\xi$ appearing in the $P_{1}$ term in eq.(\ref{eq:Smax}).
Then, steps 3 and 4 are repeated until all probabilities $P_{i}$
entering eq.(\ref{eq:Smax}) are obtained. To appreciably reduce the
residual spurious effects due to air turbulence induced by sunlight
on the top of the gallery, all the measurements were performed during
the 2017 autumn season starting at the 0 ERA hour of October 24 and
stopping at the 12 ERA hour of October 31. 

\section{RESULTS AND CONCLUSIONS}

\begin{figure}[h]
\centering{}\includegraphics[scale=0.18]{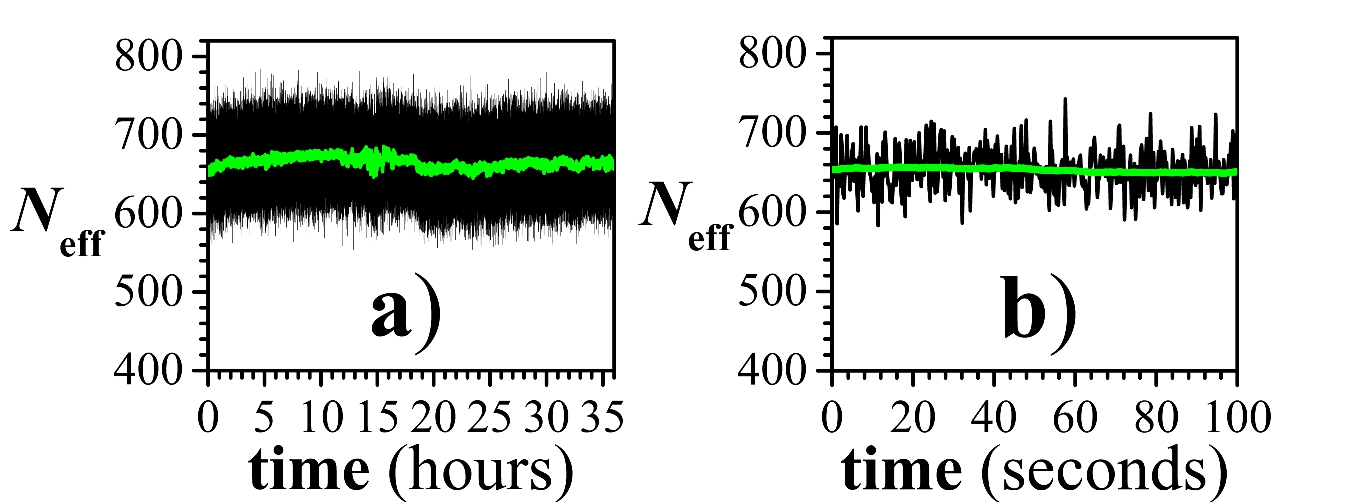}\caption{\label{fig:wide-1-1-1}a) An example of the effective coincidences
(true + spurious) versus the Greenwich \emph{ERA} time. The $2^{19}$
points are connected by black lines leading to the resulting black
region in the Figure. The acquisition time of coincidences is $\Delta t\approx0.246\,\textrm{s}$
in standard $UTC$ unities. The green full line is the result of smoothing
averaging over 200 adjacent points. The slow variations in the smoothing
curve are caused by residual noise due to sunlight on the top of the
gallery. b) A detail of the coincidences during 100 s is shown. }
\end{figure}

\begin{figure}[h]
\centering{}\includegraphics[scale=0.18]{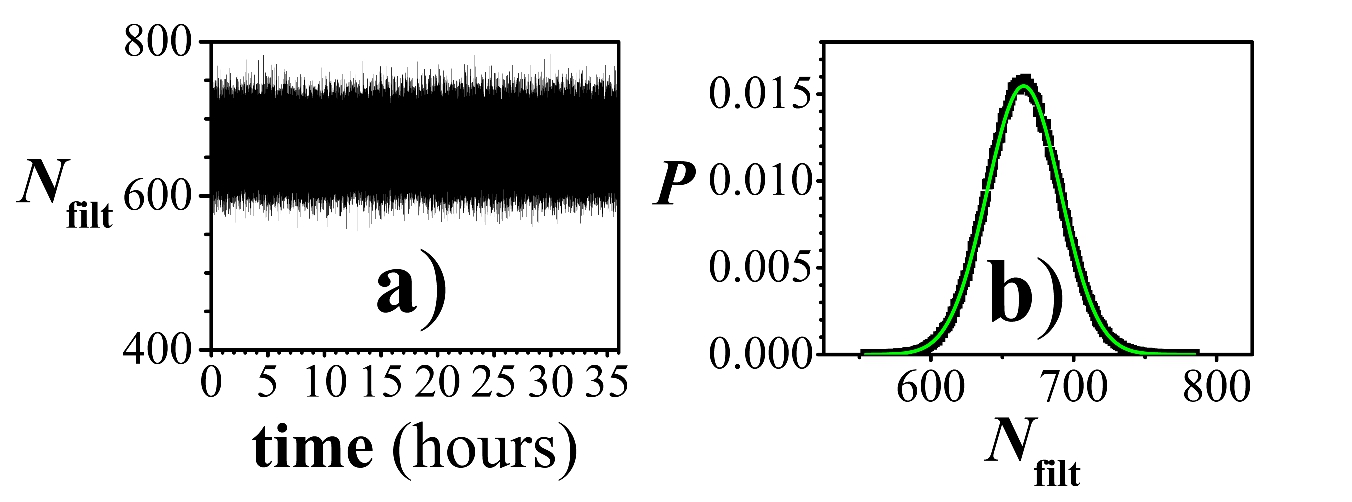}\caption{\label{fig:wide-1-1-1-1-1}a) ``Filtered'' coincidences $N_{filt}=N_{eff}-N(smoothing)+<N_{eff}>$
where the slow instrumental drift of the average value in Figure \ref{fig:wide-1-1-1}
has been subtracted. The acquisition time of coincidences is $\Delta t\approx0.246\,\textrm{s}$
and the total number of acquisitions is $2^{19}$. b) probability
distribution of the coincidences (black points). The full green curve
does not represents a best fit but it is the normal distribution predicted
by the statistic theory of counts having $\sigma^{2}=<N_{filt}>=665.042$
with no free parameters.}
\end{figure}

\begin{figure}[h]
\centering{}\includegraphics[scale=0.18]{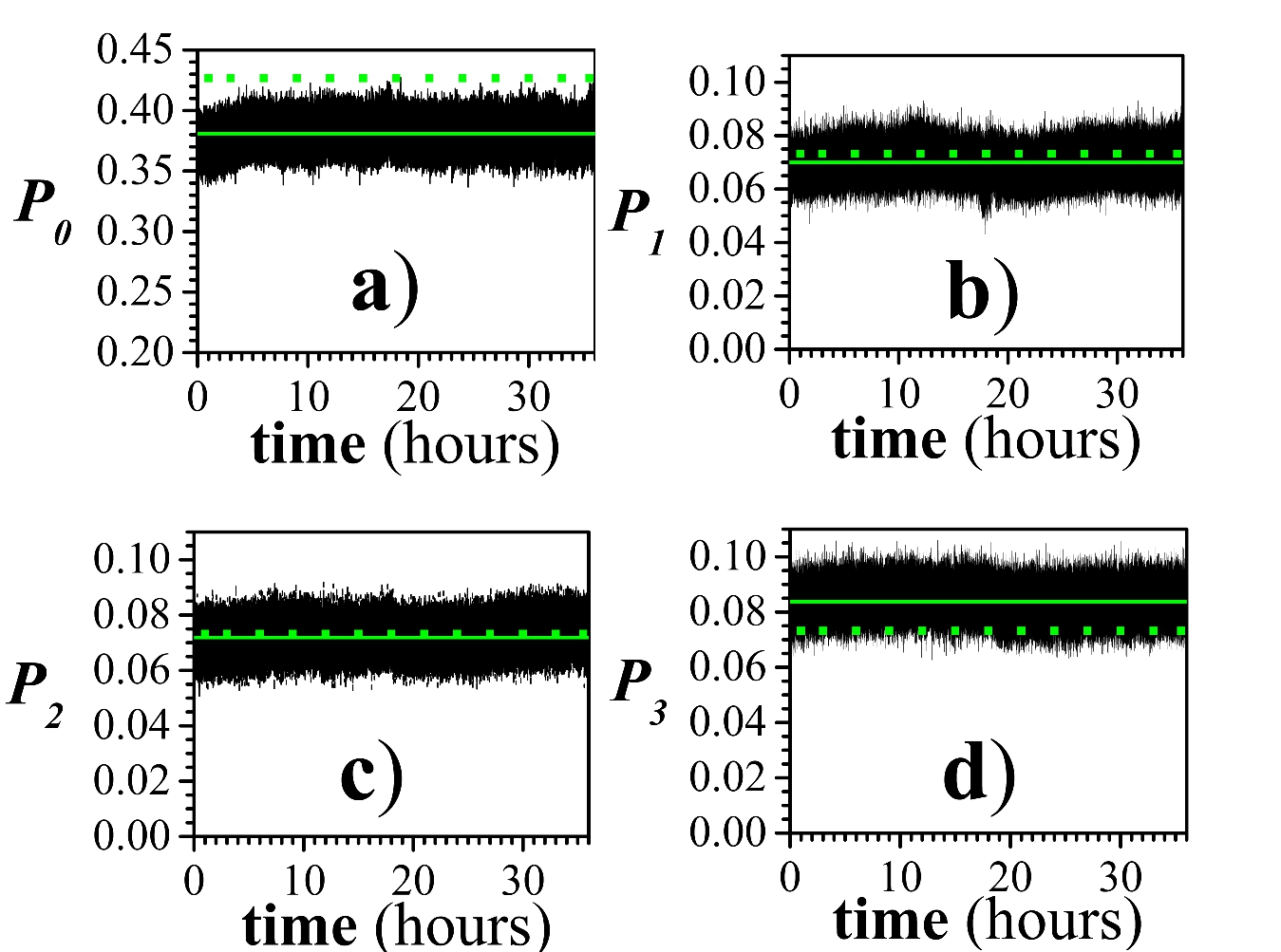}\caption{\label{fig:wide-1-1-1-2}Probabilities $P_{0},\,P_{1},\,P_{2}\,\mathrm{and\:}P_{3}$
measured in successive runs versus the Greenwich \emph{ERA} time.
The $2^{19}$ measured values are connected by straight lines leading
to the resulting black regions in the Figure. The acquisition time
is $\Delta t\approx0.246\,\textrm{s}$. The green full lines represent
the average values of the measured probabilities: $<P_{0}>=0.38087$,
$<P_{1}>=0.06999$, $<P_{2}>=0.07187$, $<P_{3}>=0.08378$. The green
dotted lines correspond to the values predicted by \emph{QM} for a
pure entangled state: $P_{0}=0.4267$, $P_{1}=P_{2}=P_{3}=0.0732$.
The difference between dotted and full lines indicates that our state
is not a pure entangled state or that some instrumental noise occurs. }
\end{figure}

Figure \ref{fig:wide-1-1-1}(a) shows an example of the effective
coincidences (true+spurious) $N_{eff}$ versus the Greenwich \emph{ERA}
time during a single run. The green full line is the result of a smoothing
obtained averaging over 200 adjacent points while a detail of the
coincidences during 100 s is shown in Figure\ref{fig:wide-1-1-1}(b).
The small slow changes that are visible in the smoothing curve are
strictly related to the daily small residual displacements of the
entangled photons beams induced by sunlight. The greater contribution
to noise in our experiment is the statistical counts noise, while
the other noise sources are virtually negligible. This is evident
if we eliminate the slow fluctuations plotting the ``filtered''
coincidences $N_{filt}=N_{eff}-N(smoothing)+<N_{eff}>$. Figure \ref{fig:wide-1-1-1-1-1}(a)
shows $N_{filt}$ versus the Greenwich \emph{ERA} time whilst Figure
\ref{fig:wide-1-1-1-1-1}(b) shows the correspondent probability distribution
\emph{P} (black points). We emphasize here that the full green line
in Figure \ref{fig:wide-1-1-1-1-1}(b) is not a best fit but it is
the Normal Gaussian function with parameters $\sigma$ and $<N_{filt}>$
that are predicted by the Statistics of counts and are given by $\sigma^{2}=<N_{filt}>=665.042$.
Figures \ref{fig:wide-1-1-1-2}(a)-\ref{fig:wide-1-1-1-2}(d) show
the probabilities $P_{i}=P(\alpha_{i},\xi_{i})=\nicefrac{N(\alpha_{i},\xi_{i})}{N_{tot}}$
obtained in the successive runs where the spurious coincidences $N_{S}=N_{A}\times N_{B}\times T_{p}/\Delta t$
have been subtracted but no filtering was performed. The black region
represents the measured values, the full green line represents the
average value whilst the green dotted line represents the value predicted
by \emph{QM} for the pure entangled state in eq.(\ref{eq:1}) (fidelity
$F=1$). The discrepancy between the full and dotted lines indicates
that our entangled state is not completely pure ($F<1$) or that some
systematic noise is present. In the simplest and rough assumption
that the breakdown of quantum correlations occurs with exactly the
Earth rotation periodicity one could calculate a $S_{max}$ value
at each \emph{ERA} time by substituting the $P_{i}$ contributions
of Figures \ref{fig:wide-1-1-1-2} measured at the same \emph{ERA}
time \emph{t} during different experimental runs into the theoretical
expression of $S_{max}$ in eq.(\ref{eq:Smax}).

\begin{figure}[h]
\centering{}\includegraphics[scale=0.18]{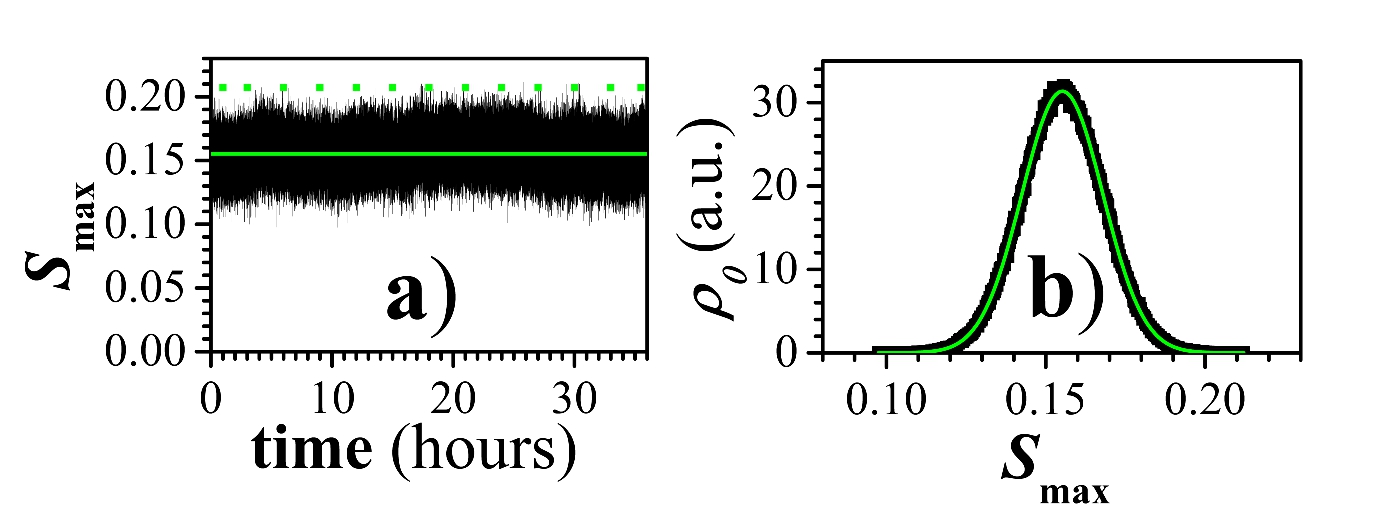}\caption{\label{fig:wide-1-1-1-3}a) $S_{max}$ versus the \emph{ERA} time
obtained using the relation $S_{max}(t)=P_{0}(t)-P_{1}(t)-P_{2}(t)-P_{3}(t)$.
The green full line is the average value $<S_{max}>=0.15523$, whilst
the green dotted line represents the \emph{QM} average value $<S_{max}>=0.207$
characterizing the pure entangled state in eq.(\ref{eq:1}). The difference
between dotted and full lines indicates that our state is not a pure
entangled state. However, the average value $<S_{max}>=0.15523$ is
sufficiently greater than zero to allow an accurate test of the Bell
inequality. b) The Frequency Distribution $\rho_{0}$ of the $2^{19}$
measured values of $S_{max}$ in arbitrary units is shown. The full
green curve is the Gaussian fit with standard deviation $\sigma=0.01272$
and $<S_{max}>=0.15523$.}
\end{figure}
With this procedure we get the results shown in Figure \ref{fig:wide-1-1-1-3}(a)
(black region) and the corresponding frequency distribution $\rho_{0}$
shown in Figure \ref{fig:wide-1-1-1-3}(b) where black points represent
the experimental results whilst the full green line is the best fit
with the Gaussian function $A\exp\left[\nicefrac{-\left(S_{max}-<S_{max}>\right)^{2}}{\left(2\sigma^{2}\right)}\right]$
with standard deviation $\sigma=0.01272$ and $<S_{max}>=0.15523$.
The green full line in Figure \ref{fig:wide-1-1-1-3}(a) shows the
average value $<S_{max}>$ whilst the green dotted line is the value
$S_{max}=$ 0.2071 predicted by \emph{QM} for the pure entangled state
in eq.(\ref{eq:1})($F=1$). No breakdown of $S_{max}$ to zero is
visible in Figure \ref{fig:wide-1-1-1-3}(a) and the lowest experimental
values of $S_{max}$ are at more then 7 standard deviations from the
maximum value $S_{max}=0$ predicted by local variables models. However,
the analysis above is not sufficient to conclude that no superluminal
effect is present. In fact, the breakdown of the \emph{QM} correlations
is predicted to occur at the two times where $\overrightarrow{\beta}\cdot\overrightarrow{AB}=0$,
where $\overrightarrow{\beta}$ is the adimensional velocity vector
of the \emph{PF }with respect to Earth. Due to the revolution motion
of the Earth around the sun and other motions (precession and nutation
of the Earth axis), vector $\overrightarrow{\beta}$ does not come
back exactly at the same orientation with respect to the Earth frame
after one \emph{ERA} day. Then, the orthogonality condition is not
satisfied exactly at the same \emph{ERA} times in different \emph{ERA}
days but some unknown time shift can occur (shifts lower than a few
min/day can be expected). Then, a rigorous test of the \emph{v}-causal
models requires a completely different analysis of the experimental
data. Denote by $t_{i1}$ and $t_{i2}$ the two unknown times during
the i-th measurement run ($i=0-3$) where the orthogonality condition
$\overrightarrow{\beta}\cdot\overrightarrow{AB}=0$ is satisfied and
by $P_{i}(t_{ij})$ with $i=0-3$ and $j=1,2$ the corresponding probabilities
measured at these times. According to the \emph{v}-causal models,
if $\beta_{t}<\beta_{t,max}$ all or someone of these probabilities
should be different from the \emph{QM} values and, thus, the correlation
parameters  

\begin{equation}
S_{max}\left(j\right)=P_{0}\left(t_{0j}\right)-\sum_{i=1}^{3}P_{i}\left(t_{ij}\right)\;,\label{eq:Smax-2}
\end{equation}
with $j=1,2$, should satisfy the Bell inequality $S_{max}\left(j\right)\leq0$
if $\beta_{t}$$<\beta_{t,max}$.

\begin{figure}[h]
\centering{}\includegraphics[scale=0.18]{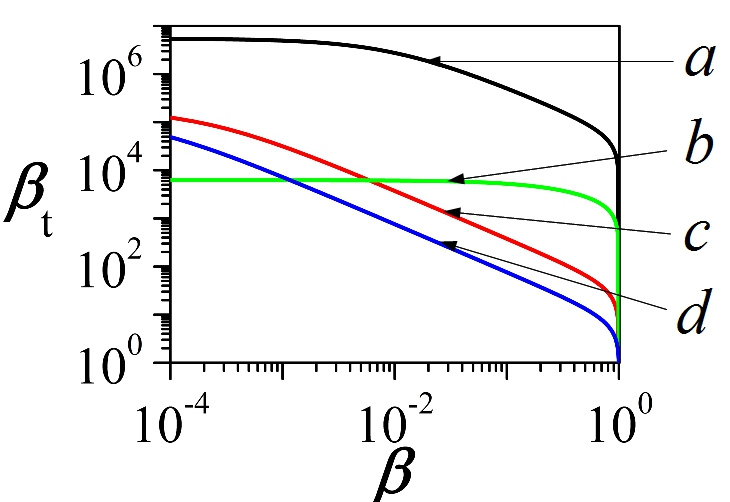}\caption{\label{fig:wide-1-1-1-3-1} Curve \emph{a} shows the $\beta_{t,max}$
values obtained in our experiment using eq.\ref{eq:betamin-2-2} ($\rho=1.83\times10^{-7}$,
$\delta t=2\,\Delta t=0.494\:{\textstyle s}$ and $\gamma=18\text{\textdegree}$)
versus the unknown adimensional  velocity $\beta$ of the \emph{PF}
for the unfavorable case $\chi=\nicefrac{\pi}{2}$; curve \emph{b}
is the result obtained in reference \citep{Cocciaro_PLA_2011} ($\rho=1.6\times10^{-4}$,
$\delta t=2\,\Delta t=8\:{\textstyle s}$ and $\gamma=0\text{\textdegree}$);
curve \emph{c} is the result obtained in reference \citep{Salart_nature_2008}
($\rho=5.4\times10^{-6}$, $\delta t=2\,\Delta t=720\:{\textstyle s}$
and $\gamma=5.9\text{\textdegree}$) and curve \emph{d} that obtained
in reference \citep{Cinesi_PhysRevLett2013} ($\rho=7.3\times10^{-6}$,
$\delta t=2\,\Delta t=3600\:{\textstyle s}$ and $\gamma=0\text{\textdegree}$).
Note that only in the case of curve \emph{d} also the locality and
the freedom-of-choice loopholes were addressed.}
\end{figure}
We do not know times $t_{ij}$ and we cannot calculate $S_{max}(j)$
but it is obvious from eq.(\ref{eq:Smax-2}) that $S_{max}\left(j\right)\geq S=MIN\left(P_{0}\right)-MAX\left(P_{1}\right)-MAX\left(P_{2}\right)-MAX\left(P_{3}\right)$
where $MIN\left(P_{i}\right)$ and $MAX\left(P_{i}\right)$ denote
the absolute minimum and maximum measured values of $P_{i}$, respectively.
From the data in Figures \ref{fig:wide-1-1-1-2} we get $S=0.04237$
and, thus, $S_{max}\left(j\right)\geq0.04237\approx3.3\:\sigma$.
This means that the probability that a value of $S_{max}\left(j\right)$
lower or equal to zero could be compatible with our measured values
is $p\leq\frac{1}{2}\mathrm{erfc}\left[\nicefrac{0.04237}{\left(\sqrt{2}\sigma\right)}\right]=4.3\cdot10^{-4}$,
where $\textrm{erfc}\left(x\right)$ is the complementary error function.
The superluminal models predict that at the least two breakdowns of
$S_{max}$ must occur in the 36 h time and, thus, the probability
that both these breakdowns happen here is $p\leq p^{2}\sim2\times10^{-7}$.
Then, we can conclude that no evidence for the presence of superluminal
communications is found and only a higher value of the lower bound
$\beta_{t,max}$ can be established. Substituting the experimental
values $\rho=1.83\times10^{-7}$ and $\delta t=2\,\Delta t=0.494\,\textrm{s}$
in eq.(\ref{eq:betamin-2-2}) one obtains $\beta_{t,max}$ as a function
of the unknown modulus $\beta$ ($\beta<1$ ) of the adimensional
velocity of the Preferred Frame\emph{ }and of his angle $\chi$ with
respect to the Earth rotation axis. We remind that eq.(\ref{eq:betamin-2-2})
holds only if angle $\chi$ is inside the interval $\left[\gamma,\pi-\gamma\right]$
where $\gamma=\pi/10$ rad, whilst $\beta_{t,max}$ sharply decreases
out of this interval \citep{Salart_nature_2008}. According to eq.(\ref{eq:betamin-2-2}),
$\beta_{t,max}$ reaches the maximum value at the borders $\chi=\gamma$
and $\chi=\pi-\gamma$ and the minimum value at $\chi=\pi/2$ . The
upper curve in Figure \ref{fig:wide-1-1-1-3-1} shows our $\beta_{t,max}$
versus the unknown adimensional velocity $\beta$ of the \emph{PF}
in the unfavorable case $\chi=\pi/2$. For \emph{PF} velocities comparable
to those of the \emph{CMB} Frame ($\beta\approx10^{-3})$ the corresponding
lower bound is $\beta_{t,max}\approx5\times10^{6}$. The lower curves
represent the experimental values of $\beta_{t,max}$ obtained in
the previous experiments \citep{Salart_nature_2008,Cocciaro_PLA_2011,Cinesi_PhysRevLett2013}.
No breakdown of quantum correlations has been observed and, thus,
we can infer that either the superluminal communications are not responsible
for quantum correlations or their adimensional velocities are greater
than $\beta_{t,max}$. Finally, it has to be noticed that it remains
open the possibility that $\beta_{t}$$<\beta_{t,max}$ but vector
$\overrightarrow{\beta}$ makes a polar angle $\chi<\gamma=\pi/10$
or $\chi>\pi-\gamma=9\pi/10$  with the Earth rotation axis.
\begin{acknowledgments}
We acknowledge the Fondazione Pisa for financial support. We acknowledge
the \emph{EGO} and \emph{VIRGO} staff that made possible the experiment
and, in particular, F. Ferrini, F. Carbognani, A. Paoli and C.Fabozzi.
A special thank to M. Bianucci (\emph{Pisa Physics Department}) and
to S. Cortese (\emph{VIRGO}) for their invaluable and continuous contribution
to the solution of a lot of electronic and informatic problems. Finally,
we acknowledge T. Faetti for his helpful suggestions on real time
procedures. 
\end{acknowledgments}

\bibliographystyle{apsrev4-1}
\bibliography{mybibPRA}

\begin{thebibliography}{32}%
\makeatletter
\providecommand \@ifxundefined [1]{%
 \@ifx{#1\undefined}
}%
\providecommand \@ifnum [1]{%
 \ifnum #1\expandafter \@firstoftwo
 \else \expandafter \@secondoftwo
 \fi
}%
\providecommand \@ifx [1]{%
 \ifx #1\expandafter \@firstoftwo
 \else \expandafter \@secondoftwo
 \fi
}%
\providecommand \natexlab [1]{#1}%
\providecommand \enquote  [1]{``#1''}%
\providecommand \bibnamefont  [1]{#1}%
\providecommand \bibfnamefont [1]{#1}%
\providecommand \citenamefont [1]{#1}%
\providecommand \href@noop [0]{\@secondoftwo}%
\providecommand \href [0]{\begingroup \@sanitize@url \@href}%
\providecommand \@href[1]{\@@startlink{#1}\@@href}%
\providecommand \@@href[1]{\endgroup#1\@@endlink}%
\providecommand \@sanitize@url [0]{\catcode `\\12\catcode `\$12\catcode
  `\&12\catcode `\#12\catcode `\^12\catcode `\_12\catcode `\%12\relax}%
\providecommand \@@startlink[1]{}%
\providecommand \@@endlink[0]{}%
\providecommand \url  [0]{\begingroup\@sanitize@url \@url }%
\providecommand \@url [1]{\endgroup\@href {#1}{\urlprefix }}%
\providecommand \urlprefix  [0]{URL }%
\providecommand \Eprint [0]{\href }%
\providecommand \doibase [0]{http://dx.doi.org/}%
\providecommand \selectlanguage [0]{\@gobble}%
\providecommand \bibinfo  [0]{\@secondoftwo}%
\providecommand \bibfield  [0]{\@secondoftwo}%
\providecommand \translation [1]{[#1]}%
\providecommand \BibitemOpen [0]{}%
\providecommand \bibitemStop [0]{}%
\providecommand \bibitemNoStop [0]{.\EOS\space}%
\providecommand \EOS [0]{\spacefactor3000\relax}%
\providecommand \BibitemShut  [1]{\csname bibitem#1\endcsname}%
\let\auto@bib@innerbib\@empty
\bibitem [{\citenamefont {Einstein}\ \emph {et~al.}(1935)\citenamefont
  {Einstein}, \citenamefont {Podolsky},\ and\ \citenamefont {Rosen}}]{EPR}%
  \BibitemOpen
  \bibfield  {author} {\bibinfo {author} {\bibfnamefont {A.}~\bibnamefont
  {Einstein}}, \bibinfo {author} {\bibfnamefont {B.}~\bibnamefont {Podolsky}},
  \ and\ \bibinfo {author} {\bibfnamefont {N.}~\bibnamefont {Rosen}},\ }\href
  {\doibase 10.1103/PhysRev.47.777} {\bibfield  {journal} {\bibinfo  {journal}
  {Phys. Rev.}\ }\textbf {\bibinfo {volume} {47}},\ \bibinfo {pages} {777}
  (\bibinfo {year} {1935})}\BibitemShut {NoStop}%
\bibitem [{\citenamefont {Bell}(1964)}]{Bell}%
  \BibitemOpen
  \bibfield  {author} {\bibinfo {author} {\bibfnamefont {J.~S.}\ \bibnamefont
  {Bell}},\ }\href@noop {} {\bibfield  {journal} {\bibinfo  {journal}
  {Physics}\ }\textbf {\bibinfo {volume} {1}},\ \bibinfo {pages} {195}
  (\bibinfo {year} {1964})}\BibitemShut {NoStop}%
\bibitem [{\citenamefont {Clauser}\ \emph {et~al.}(1969)\citenamefont
  {Clauser}, \citenamefont {Horne}, \citenamefont {A.Shimony},\ and\
  \citenamefont {Holt}}]{Clauser1_PhysRevLett1969}%
  \BibitemOpen
  \bibfield  {author} {\bibinfo {author} {\bibfnamefont {J.~F.}\ \bibnamefont
  {Clauser}}, \bibinfo {author} {\bibfnamefont {M.~A.}\ \bibnamefont {Horne}},
  \bibinfo {author} {\bibnamefont {A.Shimony}}, \ and\ \bibinfo {author}
  {\bibfnamefont {R.}~\bibnamefont {Holt}},\ }\href {\doibase
  10.1103/PhysRevLett.119.260501} {\bibfield  {journal} {\bibinfo  {journal}
  {Phys. Rev. Lett.}\ }\textbf {\bibinfo {volume} {23}},\ \bibinfo {pages}
  {880} (\bibinfo {year} {1969})}\BibitemShut {NoStop}%
\bibitem [{\citenamefont {Clauser}\ and\ \citenamefont
  {Horne}(1974)}]{Clauser2_PhysRevD1974}%
  \BibitemOpen
  \bibfield  {author} {\bibinfo {author} {\bibfnamefont {J.~F.}\ \bibnamefont
  {Clauser}}\ and\ \bibinfo {author} {\bibfnamefont {M.~A.}\ \bibnamefont
  {Horne}},\ }\href {\doibase https://doi.org/10.1103/PhysRevD.10.526}
  {\bibfield  {journal} {\bibinfo  {journal} {Phys. Rev. D}\ }\textbf {\bibinfo
  {volume} {10}},\ \bibinfo {pages} {526} (\bibinfo {year} {1974})}\BibitemShut
  {NoStop}%
\bibitem [{\citenamefont {Aspect}\ \emph {et~al.}(1982)\citenamefont {Aspect},
  \citenamefont {Dalibard},\ and\ \citenamefont {Roger}}]{Aspect}%
  \BibitemOpen
  \bibfield  {author} {\bibinfo {author} {\bibfnamefont {A.}~\bibnamefont
  {Aspect}}, \bibinfo {author} {\bibfnamefont {J.}~\bibnamefont {Dalibard}}, \
  and\ \bibinfo {author} {\bibfnamefont {G.}~\bibnamefont {Roger}},\ }\href
  {\doibase 10.1103/PhysRevLett.49.1804} {\bibfield  {journal} {\bibinfo
  {journal} {Phys. Rev. Lett.}\ }\textbf {\bibinfo {volume} {49}},\ \bibinfo
  {pages} {1804} (\bibinfo {year} {1982})}\BibitemShut {NoStop}%
\bibitem [{\citenamefont {Hensen}\ \emph {et~al.}(2015)\citenamefont {Hensen},
  \citenamefont {Bernien}, \citenamefont {Dr{\'e}au}, \citenamefont {Reiserer},
  \citenamefont {Kalb}, \citenamefont {Blok}, \citenamefont {Ruitenberg},
  \citenamefont {Vermeulen}, \citenamefont {Schouten}, \citenamefont
  {Abell{\'a}n}, \citenamefont {Amaya}, \citenamefont {Pruneri}, \citenamefont
  {Mitchell}, \citenamefont {Markham}, \citenamefont {Twitchen}, \citenamefont
  {Elkouss}, \citenamefont {Wehner}, \citenamefont {Taminiau},\ and\
  \citenamefont {Hanson}}]{Loophole1}%
  \BibitemOpen
  \bibfield  {author} {\bibinfo {author} {\bibfnamefont {B.}~\bibnamefont
  {Hensen}}, \bibinfo {author} {\bibfnamefont {H.}~\bibnamefont {Bernien}},
  \bibinfo {author} {\bibfnamefont {A.~E.}\ \bibnamefont {Dr{\'e}au}}, \bibinfo
  {author} {\bibfnamefont {A.}~\bibnamefont {Reiserer}}, \bibinfo {author}
  {\bibfnamefont {N.}~\bibnamefont {Kalb}}, \bibinfo {author} {\bibfnamefont
  {M.~S.}\ \bibnamefont {Blok}}, \bibinfo {author} {\bibfnamefont
  {J.}~\bibnamefont {Ruitenberg}}, \bibinfo {author} {\bibfnamefont {R.~F.~L.}\
  \bibnamefont {Vermeulen}}, \bibinfo {author} {\bibfnamefont {R.~N.}\
  \bibnamefont {Schouten}}, \bibinfo {author} {\bibfnamefont {C.}~\bibnamefont
  {Abell{\'a}n}}, \bibinfo {author} {\bibfnamefont {W.}~\bibnamefont {Amaya}},
  \bibinfo {author} {\bibfnamefont {V.}~\bibnamefont {Pruneri}}, \bibinfo
  {author} {\bibfnamefont {M.~W.}\ \bibnamefont {Mitchell}}, \bibinfo {author}
  {\bibfnamefont {M.}~\bibnamefont {Markham}}, \bibinfo {author} {\bibfnamefont
  {D.~J.}\ \bibnamefont {Twitchen}}, \bibinfo {author} {\bibfnamefont
  {D.}~\bibnamefont {Elkouss}}, \bibinfo {author} {\bibfnamefont
  {S.}~\bibnamefont {Wehner}}, \bibinfo {author} {\bibfnamefont {T.~H.}\
  \bibnamefont {Taminiau}}, \ and\ \bibinfo {author} {\bibfnamefont
  {R.}~\bibnamefont {Hanson}},\ }\href {http://dx.doi.org/10.1038/nature15759}
  {\bibfield  {journal} {\bibinfo  {journal} {Nature}\ }\textbf {\bibinfo
  {volume} {526}},\ \bibinfo {pages} {682 EP } (\bibinfo {year}
  {2015})}\BibitemShut {NoStop}%
\bibitem [{\citenamefont {Giustina}\ \emph {et~al.}(2015)\citenamefont
  {Giustina}, \citenamefont {Versteegh}, \citenamefont {Wengerowsky},
  \citenamefont {Handsteiner}, \citenamefont {Hochrainer}, \citenamefont
  {Phelan}, \citenamefont {Steinlechner}, \citenamefont {Kofler}, \citenamefont
  {Larsson}, \citenamefont {Abell\'an}, \citenamefont {Amaya}, \citenamefont
  {Pruneri}, \citenamefont {Mitchell}, \citenamefont {Beyer}, \citenamefont
  {Gerrits}, \citenamefont {Lita}, \citenamefont {Shalm}, \citenamefont {Nam},
  \citenamefont {Scheidl}, \citenamefont {Ursin}, \citenamefont {Wittmann},\
  and\ \citenamefont {Zeilinger}}]{Loophole2}%
  \BibitemOpen
  \bibfield  {author} {\bibinfo {author} {\bibfnamefont {M.}~\bibnamefont
  {Giustina}}, \bibinfo {author} {\bibfnamefont {M.~A.~M.}\ \bibnamefont
  {Versteegh}}, \bibinfo {author} {\bibfnamefont {S.}~\bibnamefont
  {Wengerowsky}}, \bibinfo {author} {\bibfnamefont {J.}~\bibnamefont
  {Handsteiner}}, \bibinfo {author} {\bibfnamefont {A.}~\bibnamefont
  {Hochrainer}}, \bibinfo {author} {\bibfnamefont {K.}~\bibnamefont {Phelan}},
  \bibinfo {author} {\bibfnamefont {F.}~\bibnamefont {Steinlechner}}, \bibinfo
  {author} {\bibfnamefont {J.}~\bibnamefont {Kofler}}, \bibinfo {author}
  {\bibfnamefont {J.-A.}\ \bibnamefont {Larsson}}, \bibinfo {author}
  {\bibfnamefont {C.}~\bibnamefont {Abell\'an}}, \bibinfo {author}
  {\bibfnamefont {W.}~\bibnamefont {Amaya}}, \bibinfo {author} {\bibfnamefont
  {V.}~\bibnamefont {Pruneri}}, \bibinfo {author} {\bibfnamefont {M.~W.}\
  \bibnamefont {Mitchell}}, \bibinfo {author} {\bibfnamefont {J.}~\bibnamefont
  {Beyer}}, \bibinfo {author} {\bibfnamefont {T.}~\bibnamefont {Gerrits}},
  \bibinfo {author} {\bibfnamefont {A.~E.}\ \bibnamefont {Lita}}, \bibinfo
  {author} {\bibfnamefont {L.~K.}\ \bibnamefont {Shalm}}, \bibinfo {author}
  {\bibfnamefont {S.~W.}\ \bibnamefont {Nam}}, \bibinfo {author} {\bibfnamefont
  {T.}~\bibnamefont {Scheidl}}, \bibinfo {author} {\bibfnamefont
  {R.}~\bibnamefont {Ursin}}, \bibinfo {author} {\bibfnamefont
  {B.}~\bibnamefont {Wittmann}}, \ and\ \bibinfo {author} {\bibfnamefont
  {A.}~\bibnamefont {Zeilinger}},\ }\href {\doibase
  10.1103/PhysRevLett.115.250401} {\bibfield  {journal} {\bibinfo  {journal}
  {Phys. Rev. Lett.}\ }\textbf {\bibinfo {volume} {115}},\ \bibinfo {pages}
  {250401} (\bibinfo {year} {2015})}\BibitemShut {NoStop}%
\bibitem [{\citenamefont {Shalm}\ \emph {et~al.}(2015)\citenamefont {Shalm},
  \citenamefont {Meyer-Scott}, \citenamefont {Christensen}, \citenamefont
  {Bierhorst}, \citenamefont {Wayne}, \citenamefont {Stevens}, \citenamefont
  {Gerrits}, \citenamefont {Glancy}, \citenamefont {Hamel}, \citenamefont
  {Allman}, \citenamefont {Coakley}, \citenamefont {Dyer}, \citenamefont
  {Hodge}, \citenamefont {Lita}, \citenamefont {Verma}, \citenamefont
  {Lambrocco}, \citenamefont {Tortorici}, \citenamefont {Migdall},
  \citenamefont {Zhang}, \citenamefont {Kumor}, \citenamefont {Farr},
  \citenamefont {Marsili}, \citenamefont {Shaw}, \citenamefont {Stern},
  \citenamefont {Abell\'an}, \citenamefont {Amaya}, \citenamefont {Pruneri},
  \citenamefont {Jennewein}, \citenamefont {Mitchell}, \citenamefont {Kwiat},
  \citenamefont {Bienfang}, \citenamefont {Mirin}, \citenamefont {Knill},\ and\
  \citenamefont {Nam}}]{Loophole3}%
  \BibitemOpen
  \bibfield  {author} {\bibinfo {author} {\bibfnamefont {L.~K.}\ \bibnamefont
  {Shalm}}, \bibinfo {author} {\bibfnamefont {E.}~\bibnamefont {Meyer-Scott}},
  \bibinfo {author} {\bibfnamefont {B.~G.}\ \bibnamefont {Christensen}},
  \bibinfo {author} {\bibfnamefont {P.}~\bibnamefont {Bierhorst}}, \bibinfo
  {author} {\bibfnamefont {M.~A.}\ \bibnamefont {Wayne}}, \bibinfo {author}
  {\bibfnamefont {M.~J.}\ \bibnamefont {Stevens}}, \bibinfo {author}
  {\bibfnamefont {T.}~\bibnamefont {Gerrits}}, \bibinfo {author} {\bibfnamefont
  {S.}~\bibnamefont {Glancy}}, \bibinfo {author} {\bibfnamefont {D.~R.}\
  \bibnamefont {Hamel}}, \bibinfo {author} {\bibfnamefont {M.~S.}\ \bibnamefont
  {Allman}}, \bibinfo {author} {\bibfnamefont {K.~J.}\ \bibnamefont {Coakley}},
  \bibinfo {author} {\bibfnamefont {S.~D.}\ \bibnamefont {Dyer}}, \bibinfo
  {author} {\bibfnamefont {C.}~\bibnamefont {Hodge}}, \bibinfo {author}
  {\bibfnamefont {A.~E.}\ \bibnamefont {Lita}}, \bibinfo {author}
  {\bibfnamefont {V.~B.}\ \bibnamefont {Verma}}, \bibinfo {author}
  {\bibfnamefont {C.}~\bibnamefont {Lambrocco}}, \bibinfo {author}
  {\bibfnamefont {E.}~\bibnamefont {Tortorici}}, \bibinfo {author}
  {\bibfnamefont {A.~L.}\ \bibnamefont {Migdall}}, \bibinfo {author}
  {\bibfnamefont {Y.}~\bibnamefont {Zhang}}, \bibinfo {author} {\bibfnamefont
  {D.~R.}\ \bibnamefont {Kumor}}, \bibinfo {author} {\bibfnamefont {W.~H.}\
  \bibnamefont {Farr}}, \bibinfo {author} {\bibfnamefont {F.}~\bibnamefont
  {Marsili}}, \bibinfo {author} {\bibfnamefont {M.~D.}\ \bibnamefont {Shaw}},
  \bibinfo {author} {\bibfnamefont {J.~A.}\ \bibnamefont {Stern}}, \bibinfo
  {author} {\bibfnamefont {C.}~\bibnamefont {Abell\'an}}, \bibinfo {author}
  {\bibfnamefont {W.}~\bibnamefont {Amaya}}, \bibinfo {author} {\bibfnamefont
  {V.}~\bibnamefont {Pruneri}}, \bibinfo {author} {\bibfnamefont
  {T.}~\bibnamefont {Jennewein}}, \bibinfo {author} {\bibfnamefont {M.~W.}\
  \bibnamefont {Mitchell}}, \bibinfo {author} {\bibfnamefont {P.~G.}\
  \bibnamefont {Kwiat}}, \bibinfo {author} {\bibfnamefont {J.~C.}\ \bibnamefont
  {Bienfang}}, \bibinfo {author} {\bibfnamefont {R.~P.}\ \bibnamefont {Mirin}},
  \bibinfo {author} {\bibfnamefont {E.}~\bibnamefont {Knill}}, \ and\ \bibinfo
  {author} {\bibfnamefont {S.~W.}\ \bibnamefont {Nam}},\ }\href {\doibase
  10.1103/PhysRevLett.115.250402} {\bibfield  {journal} {\bibinfo  {journal}
  {Phys. Rev. Lett.}\ }\textbf {\bibinfo {volume} {115}},\ \bibinfo {pages}
  {250402} (\bibinfo {year} {2015})}\BibitemShut {NoStop}%
\bibitem [{\citenamefont {Rosenfeld}\ \emph {et~al.}(2017)\citenamefont
  {Rosenfeld}, \citenamefont {Burchardt}, \citenamefont {Garthoff},
  \citenamefont {Redeker}, \citenamefont {Ortegel}, \citenamefont {Rau},\ and\
  \citenamefont {Weinfurter}}]{Loophole4}%
  \BibitemOpen
  \bibfield  {author} {\bibinfo {author} {\bibfnamefont {W.}~\bibnamefont
  {Rosenfeld}}, \bibinfo {author} {\bibfnamefont {D.}~\bibnamefont
  {Burchardt}}, \bibinfo {author} {\bibfnamefont {R.}~\bibnamefont {Garthoff}},
  \bibinfo {author} {\bibfnamefont {K.}~\bibnamefont {Redeker}}, \bibinfo
  {author} {\bibfnamefont {N.}~\bibnamefont {Ortegel}}, \bibinfo {author}
  {\bibfnamefont {M.}~\bibnamefont {Rau}}, \ and\ \bibinfo {author}
  {\bibfnamefont {H.}~\bibnamefont {Weinfurter}},\ }\href {\doibase
  10.1103/PhysRevLett.119.010402} {\bibfield  {journal} {\bibinfo  {journal}
  {Phys. Rev. Lett.}\ }\textbf {\bibinfo {volume} {119}},\ \bibinfo {pages}
  {010402} (\bibinfo {year} {2017})}\BibitemShut {NoStop}%
\bibitem [{\citenamefont {Eberhard}(1989)}]{Eberhard_1989}%
  \BibitemOpen
  \bibfield  {author} {\bibinfo {author} {\bibfnamefont {P.~H.}\ \bibnamefont
  {Eberhard}},\ }in\ \href@noop {} {\emph {\bibinfo {booktitle} {Quantum theory
  and pictures of reality: foundations, interpretations, and new aspects}}},\
  \bibinfo {editor} {edited by\ \bibinfo {editor} {\bibfnamefont
  {W.}~\bibnamefont {Schommers}}}\ (\bibinfo  {publisher} {{Springer-Verlag}},\
  \bibinfo {address} {Berlin; New York},\ \bibinfo {year} {1989})\BibitemShut
  {NoStop}%
\bibitem [{\citenamefont {Bohm}\ and\ \citenamefont
  {Hiley}(1993)}]{Bohm_undivided_1993}%
  \BibitemOpen
  \bibfield  {author} {\bibinfo {author} {\bibfnamefont {D.}~\bibnamefont
  {Bohm}}\ and\ \bibinfo {author} {\bibfnamefont {B.~J.}\ \bibnamefont
  {Hiley}},\ }\href@noop {} {\emph {\bibinfo {title} {The undivided universe:
  an ontological interpretation of quantum mechanics}}}\ (\bibinfo  {publisher}
  {Routledge,London,},\ \bibinfo {year} {1993})\BibitemShut {NoStop}%
\bibitem [{Note1()}]{Note1}%
  \BibitemOpen
  \bibinfo {note} {The key idea is that, in a typical \protect \emph {EPR}
  experiment, the two measurements are not exactly simultaneous. When the first
  measurement is performed in the point \protect \emph {A}, a collapsing wave
  propagates superluminally in the space starting from \protect \emph {A}. The
  predicted \protect \emph {QM} correlations (e. g. the violation of the Bell
  inequality) occur only if the collapsing wave gets the second particle before
  its measurement.}\BibitemShut {Stop}%
\bibitem [{\citenamefont {Gisin}(2014)}]{Gisin2014}%
  \BibitemOpen
  \bibfield  {author} {\bibinfo {author} {\bibfnamefont {N.}~\bibnamefont
  {Gisin}},\ }\enquote {\bibinfo {title} {Quantum correlations in newtonian
  space and time: Faster than light communication or nonlocality},}\ in\ \href
  {\doibase 10.1007/978-88-470-5217-8_12} {\emph {\bibinfo {booktitle} {Quantum
  Theory: A Two-Time Success Story: Yakir Aharonov Festschrift}}},\ \bibinfo
  {editor} {edited by\ \bibinfo {editor} {\bibfnamefont {D.~C.}\ \bibnamefont
  {Struppa}}\ and\ \bibinfo {editor} {\bibfnamefont {J.~M.}\ \bibnamefont
  {Tollaksen}}}\ (\bibinfo  {publisher} {Springer Milan},\ \bibinfo {address}
  {Milano},\ \bibinfo {year} {2014})\ pp.\ \bibinfo {pages}
  {185--203}\BibitemShut {NoStop}%
\bibitem [{\citenamefont {Cocciaro}(2013)}]{Cocciaro_2013_ShutYourselfUp}%
  \BibitemOpen
  \bibfield  {author} {\bibinfo {author} {\bibfnamefont {B.}~\bibnamefont
  {Cocciaro}},\ }\href {\doibase 10.4006/0836-1398-26.4.531} {\bibfield
  {journal} {\bibinfo  {journal} {Physics Essays}\ }\textbf {\bibinfo {volume}
  {26}},\ \bibinfo {pages} {531} (\bibinfo {year} {2013})}\BibitemShut
  {NoStop}%
\bibitem [{\citenamefont {Cocciaro}(2015)}]{Cocciaro3_DICE2015}%
  \BibitemOpen
  \bibfield  {author} {\bibinfo {author} {\bibfnamefont {B.}~\bibnamefont
  {Cocciaro}},\ }\href {\doibase 10.1088/1742-6596/626/1/012054} {\bibfield
  {journal} {\bibinfo  {journal} {Journal of Physics: Conference Series}\
  }\textbf {\bibinfo {volume} {626}},\ \bibinfo {pages} {012054} (\bibinfo
  {year} {2015})}\BibitemShut {NoStop}%
\bibitem [{\citenamefont {Bancal}\ \emph {et~al.}(2012)\citenamefont {Bancal},
  \citenamefont {Pironio}, \citenamefont {Ac\ifmmode~\acute{i}\else
  \'{i}\fi{}n}, \citenamefont {Liang}, \citenamefont {Scarani},\ and\
  \citenamefont {Gisin}}]{Bancal_NatPhys_2012}%
  \BibitemOpen
  \bibfield  {author} {\bibinfo {author} {\bibfnamefont {J.}~\bibnamefont
  {Bancal}}, \bibinfo {author} {\bibfnamefont {S.}~\bibnamefont {Pironio}},
  \bibinfo {author} {\bibfnamefont {A.}~\bibnamefont {Ac\ifmmode~\acute{i}\else
  \'{i}\fi{}n}}, \bibinfo {author} {\bibfnamefont {Y.}~\bibnamefont {Liang}},
  \bibinfo {author} {\bibfnamefont {V.}~\bibnamefont {Scarani}}, \ and\
  \bibinfo {author} {\bibfnamefont {N.}~\bibnamefont {Gisin}},\ }\href
  {\doibase 10.1038/nphys2460} {\bibfield  {journal} {\bibinfo  {journal} {Nat.
  Phys.}\ }\textbf {\bibinfo {volume} {8}},\ \bibinfo {pages} {867} (\bibinfo
  {year} {2012})}\BibitemShut {NoStop}%
\bibitem [{\citenamefont {Barnea}\ \emph {et~al.}(2013)\citenamefont {Barnea},
  \citenamefont {Bancal}, \citenamefont {Liang},\ and\ \citenamefont
  {Gisin}}]{Barnea_PhysRevA2013}%
  \BibitemOpen
  \bibfield  {author} {\bibinfo {author} {\bibfnamefont {T.~J.}\ \bibnamefont
  {Barnea}}, \bibinfo {author} {\bibfnamefont {J.-D.}\ \bibnamefont {Bancal}},
  \bibinfo {author} {\bibfnamefont {Y.-C.}\ \bibnamefont {Liang}}, \ and\
  \bibinfo {author} {\bibfnamefont {N.}~\bibnamefont {Gisin}},\ }\href
  {\doibase 10.1103/PhysRevA.88.022123} {\bibfield  {journal} {\bibinfo
  {journal} {Phys. Rev. A}\ }\textbf {\bibinfo {volume} {88}},\ \bibinfo
  {pages} {022123} (\bibinfo {year} {2013})}\BibitemShut {NoStop}%
\bibitem [{\citenamefont {Aspect}(2002)}]{Aspect_2002}%
  \BibitemOpen
  \bibfield  {author} {\bibinfo {author} {\bibfnamefont {A.}~\bibnamefont
  {Aspect}},\ }in\ \href@noop {} {\emph {\bibinfo {booktitle} {Quantum
  {(Un)speakables}: From Bell to Quantum Information}}},\ \bibinfo {editor}
  {edited by\ \bibinfo {editor} {\bibfnamefont {R.~A.}\ \bibnamefont
  {Bertlmann}}\ and\ \bibinfo {editor} {\bibfnamefont {A.}~\bibnamefont
  {Zeilinger}}}\ (\bibinfo  {publisher} {Springer},\ \bibinfo {year}
  {2002})\BibitemShut {NoStop}%
\bibitem [{\citenamefont {Cocciaro}\ \emph {et~al.}(2017)\citenamefont
  {Cocciaro}, \citenamefont {Faetti},\ and\ \citenamefont
  {Fronzoni}}]{Cocciaro2_DICE2016}%
  \BibitemOpen
  \bibfield  {author} {\bibinfo {author} {\bibfnamefont {B.}~\bibnamefont
  {Cocciaro}}, \bibinfo {author} {\bibfnamefont {S.}~\bibnamefont {Faetti}}, \
  and\ \bibinfo {author} {\bibfnamefont {L.}~\bibnamefont {Fronzoni}},\ }\href
  {\doibase 10.1088/1742-6596/880/1/012036} {\bibfield  {journal} {\bibinfo
  {journal} {Journal of Physics: Conference Series}\ }\textbf {\bibinfo
  {volume} {880}},\ \bibinfo {pages} {012036} (\bibinfo {year}
  {2017})}\BibitemShut {NoStop}%
\bibitem [{\citenamefont {Salart}\ \emph
  {et~al.}(2008{\natexlab{a}})\citenamefont {Salart}, \citenamefont {Baas},
  \citenamefont {Branciard}, \citenamefont {Gisin},\ and\ \citenamefont
  {Zbinden}}]{Salart_nature_2008}%
  \BibitemOpen
  \bibfield  {author} {\bibinfo {author} {\bibfnamefont {D.}~\bibnamefont
  {Salart}}, \bibinfo {author} {\bibfnamefont {A.}~\bibnamefont {Baas}},
  \bibinfo {author} {\bibfnamefont {C.}~\bibnamefont {Branciard}}, \bibinfo
  {author} {\bibfnamefont {N.}~\bibnamefont {Gisin}}, \ and\ \bibinfo {author}
  {\bibfnamefont {H.}~\bibnamefont {Zbinden}},\ }\href {\doibase
  10.1038/nature07121} {\bibfield  {journal} {\bibinfo  {journal} {Nature}\
  }\textbf {\bibinfo {volume} {454}},\ \bibinfo {pages} {861} (\bibinfo {year}
  {2008}{\natexlab{a}})}\BibitemShut {NoStop}%
\bibitem [{\citenamefont {Cocciaro}\ \emph {et~al.}(2011)\citenamefont
  {Cocciaro}, \citenamefont {Faetti},\ and\ \citenamefont
  {Fronzoni}}]{Cocciaro_PLA_2011}%
  \BibitemOpen
  \bibfield  {author} {\bibinfo {author} {\bibfnamefont {B.}~\bibnamefont
  {Cocciaro}}, \bibinfo {author} {\bibfnamefont {S.}~\bibnamefont {Faetti}}, \
  and\ \bibinfo {author} {\bibfnamefont {L.}~\bibnamefont {Fronzoni}},\ }\href
  {\doibase 10.1016/j.physleta.2010.10.064} {\bibfield  {journal} {\bibinfo
  {journal} {Phys. Lett. A}\ }\textbf {\bibinfo {volume} {375}},\ \bibinfo
  {pages} {379} (\bibinfo {year} {2011})}\BibitemShut {NoStop}%
\bibitem [{\citenamefont {Yin}\ \emph {et~al.}(2013)\citenamefont {Yin},
  \citenamefont {Cao}, \citenamefont {Yong}, \citenamefont {Ren}, \citenamefont
  {Liang}, \citenamefont {Liao}, \citenamefont {Zhou}, \citenamefont {Liu},
  \citenamefont {Wu}, \citenamefont {Pan}, \citenamefont {Li}, \citenamefont
  {Liu}, \citenamefont {Zhang}, \citenamefont {Peng},\ and\ \citenamefont
  {Pan}}]{Cinesi_PhysRevLett2013}%
  \BibitemOpen
  \bibfield  {author} {\bibinfo {author} {\bibfnamefont {J.}~\bibnamefont
  {Yin}}, \bibinfo {author} {\bibfnamefont {Y.}~\bibnamefont {Cao}}, \bibinfo
  {author} {\bibfnamefont {H.-L.}\ \bibnamefont {Yong}}, \bibinfo {author}
  {\bibfnamefont {J.-G.}\ \bibnamefont {Ren}}, \bibinfo {author} {\bibfnamefont
  {H.}~\bibnamefont {Liang}}, \bibinfo {author} {\bibfnamefont {S.-K.}\
  \bibnamefont {Liao}}, \bibinfo {author} {\bibfnamefont {F.}~\bibnamefont
  {Zhou}}, \bibinfo {author} {\bibfnamefont {C.}~\bibnamefont {Liu}}, \bibinfo
  {author} {\bibfnamefont {Y.-P.}\ \bibnamefont {Wu}}, \bibinfo {author}
  {\bibfnamefont {G.-S.}\ \bibnamefont {Pan}}, \bibinfo {author} {\bibfnamefont
  {L.}~\bibnamefont {Li}}, \bibinfo {author} {\bibfnamefont {N.-L.}\
  \bibnamefont {Liu}}, \bibinfo {author} {\bibfnamefont {Q.}~\bibnamefont
  {Zhang}}, \bibinfo {author} {\bibfnamefont {C.-Z.}\ \bibnamefont {Peng}}, \
  and\ \bibinfo {author} {\bibfnamefont {J.-W.}\ \bibnamefont {Pan}},\ }\href
  {\doibase 10.1103/PhysRevLett.110.260407} {\bibfield  {journal} {\bibinfo
  {journal} {Phys. Rev. Lett.}\ }\textbf {\bibinfo {volume} {110}},\ \bibinfo
  {pages} {260407} (\bibinfo {year} {2013})}\BibitemShut {NoStop}%
\bibitem [{\citenamefont {Salart}\ \emph
  {et~al.}(2008{\natexlab{b}})\citenamefont {Salart}, \citenamefont {Baas},
  \citenamefont {Branciard}, \citenamefont {Gisin},\ and\ \citenamefont
  {Zbinden}}]{SalartReply}%
  \BibitemOpen
  \bibfield  {author} {\bibinfo {author} {\bibfnamefont {D.}~\bibnamefont
  {Salart}}, \bibinfo {author} {\bibfnamefont {A.}~\bibnamefont {Baas}},
  \bibinfo {author} {\bibfnamefont {C.}~\bibnamefont {Branciard}}, \bibinfo
  {author} {\bibfnamefont {N.}~\bibnamefont {Gisin}}, \ and\ \bibinfo {author}
  {\bibfnamefont {H.}~\bibnamefont {Zbinden}},\ }\href
  {http://arxiv.org/abs/0810.4607} {\bibfield  {journal} {\bibinfo  {journal}
  {arXiv:0810.4607 [quant-ph]}\ } (\bibinfo {year}
  {2008}{\natexlab{b}})}\BibitemShut {NoStop}%
\bibitem [{\citenamefont {{European Gravitational
  Observatory,https://www.ego-gw.it/}}()}]{EGO}%
  \BibitemOpen
  \bibfield  {author} {\bibinfo {author} {\bibnamefont {{European Gravitational
  Observatory,https://www.ego-gw.it/}}},\ }\href {https://www.ego-gw.it/}
  {}\BibitemShut {NoStop}%
\bibitem [{\citenamefont {Altepeter}\ \emph {et~al.}(2005)\citenamefont
  {Altepeter}, \citenamefont {Jeffrey},\ and\ \citenamefont
  {Kwiat}}]{Kwiat_OptExpr_2005}%
  \BibitemOpen
  \bibfield  {author} {\bibinfo {author} {\bibfnamefont {J.}~\bibnamefont
  {Altepeter}}, \bibinfo {author} {\bibfnamefont {E.}~\bibnamefont {Jeffrey}},
  \ and\ \bibinfo {author} {\bibfnamefont {P.}~\bibnamefont {Kwiat}},\ }\href
  {\doibase 10.1364/OPEX.13.008951} {\bibfield  {journal} {\bibinfo  {journal}
  {Opt. Express}\ }\textbf {\bibinfo {volume} {13}},\ \bibinfo {pages} {8951}
  (\bibinfo {year} {2005})}\BibitemShut {NoStop}%
\bibitem [{\citenamefont {Akselrod}\ \emph {et~al.}(2007)\citenamefont
  {Akselrod}, \citenamefont {Altepeter}, \citenamefont {Jeffrey},\ and\
  \citenamefont {Kwiat}}]{Kwiat_OptExpr_2007}%
  \BibitemOpen
  \bibfield  {author} {\bibinfo {author} {\bibfnamefont {G.~M.}\ \bibnamefont
  {Akselrod}}, \bibinfo {author} {\bibfnamefont {J.~B.}\ \bibnamefont
  {Altepeter}}, \bibinfo {author} {\bibfnamefont {E.~R.}\ \bibnamefont
  {Jeffrey}}, \ and\ \bibinfo {author} {\bibfnamefont {P.~G.}\ \bibnamefont
  {Kwiat}},\ }\href {\doibase 10.1364/OE.15.005260} {\bibfield  {journal}
  {\bibinfo  {journal} {Opt. Express}\ }\textbf {\bibinfo {volume} {15}},\
  \bibinfo {pages} {5260} (\bibinfo {year} {2007})}\BibitemShut {NoStop}%
\bibitem [{\citenamefont {Rangarajan}\ \emph {et~al.}(2009)\citenamefont
  {Rangarajan}, \citenamefont {Goggin},\ and\ \citenamefont
  {Kwiat}}]{Kwiat_OptExpr_2009}%
  \BibitemOpen
  \bibfield  {author} {\bibinfo {author} {\bibfnamefont {R.}~\bibnamefont
  {Rangarajan}}, \bibinfo {author} {\bibfnamefont {M.}~\bibnamefont {Goggin}},
  \ and\ \bibinfo {author} {\bibfnamefont {P.}~\bibnamefont {Kwiat}},\ }\href
  {\doibase 10.1364/OE.17.018920} {\bibfield  {journal} {\bibinfo  {journal}
  {Opt. Express}\ }\textbf {\bibinfo {volume} {17}},\ \bibinfo {pages} {18920}
  (\bibinfo {year} {2009})}\BibitemShut {NoStop}%
\bibitem [{\citenamefont {Kwiat}\ \emph {et~al.}(1999)\citenamefont {Kwiat},
  \citenamefont {Waks}, \citenamefont {White}, \citenamefont {Appelbaum},\ and\
  \citenamefont {Eberhard}}]{Kwiat_PhysRevA_1999}%
  \BibitemOpen
  \bibfield  {author} {\bibinfo {author} {\bibfnamefont {P.~G.}\ \bibnamefont
  {Kwiat}}, \bibinfo {author} {\bibfnamefont {E.}~\bibnamefont {Waks}},
  \bibinfo {author} {\bibfnamefont {A.~G.}\ \bibnamefont {White}}, \bibinfo
  {author} {\bibfnamefont {I.}~\bibnamefont {Appelbaum}}, \ and\ \bibinfo
  {author} {\bibfnamefont {P.~H.}\ \bibnamefont {Eberhard}},\ }\href {\doibase
  10.1103/PhysRevA.60.R773} {\bibfield  {journal} {\bibinfo  {journal} {Phys.
  Rev. A}\ }\textbf {\bibinfo {volume} {60}},\ \bibinfo {pages} {R773}
  (\bibinfo {year} {1999})}\BibitemShut {NoStop}%
\bibitem [{\citenamefont {P.K.~Seidelmann}(1992)}]{UTC}%
  \BibitemOpen
  \bibfield  {author} {\bibinfo {author} {\bibfnamefont {L.~D.}\ \bibnamefont
  {P.K.~Seidelmann}, \bibfnamefont {B.~Guinot}},\ }\href@noop {} {\emph
  {\bibinfo {title} {Explanatory Supplement to the Astronomical
  Almanac,cp.2}}}\ (\bibinfo  {publisher} {US Naval Observatory,University
  Science books,Mill Walley,CA},\ \bibinfo {year} {1992})\BibitemShut {NoStop}%
\bibitem [{UT1()}]{UT1time}%
  \BibitemOpen
  \href@noop {} {}\bibinfo {note} {{National Radio Astronomy Observatory, Rick
  Fisher homepage,
  \url{https://www.cv.nrao.edu/~rfisher/Ephemerides/times.htm}}}\BibitemShut
  {NoStop}%
\bibitem [{IER()}]{IERS}%
  \BibitemOpen
  \href@noop {} {}\bibinfo {note} {International Earth Rotation and Reference
  Systems Service, \url{https://datacenter.iers.org/}}\BibitemShut {NoStop}%
\bibitem [{Note2()}]{Note2}%
  \BibitemOpen
  \bibinfo {note} {Due to the microseconds quantization of the $DAQ$ clock, the
  acquisition time interval $\Delta t$ = 246517 ${\textstyle \mu s}$ is smaller
  than $T_{0}/2^{19}=246517.46170157..{\textstyle \mu s}.$ by the quantity
  $\Delta _{q}t$ = 0.46170157..$\mu s$. Then, the i-th acquisition interval is
  shifted by (i-1)$\times \Delta _{q}t$ with respect to the correct value
  (i-1)$\times \Delta t$ . As soon as this shift becomes greater than 100
  ${\textstyle \mu s}$ ( for a given $i$), the Labview program increases the
  acquisition time of the i-th interval to $\Delta t+100\protect \tmspace
  +\thinmuskip {.1667em}{\textstyle {\displaystyle {\textstyle \mu s}}}$. The
  same procedure is repeated whenever the successive shifts just exceeds the
  100 ${\textstyle \mu s}$ value. In such a way the maximum residual shifts are
  always lower than $\approx 100\protect \tmspace +\thinmuskip
  {.1667em}{\textstyle {\displaystyle {\textstyle \mu s}}}$ and, thus, are
  negligible with respect to the width $\Delta t$ = 246517 ${\textstyle \mu s}$
  of each acquisition interval.}\BibitemShut {Stop}%
\end{thebibliography}%

\end{document}